\documentclass[12pt,preprint]{emulateapj}

\usepackage{color}
\usepackage{url}

\usepackage{amsmath}

\newcommand{\red}{\textcolor{black}}

\newcommand{\average}[1]{\ensuremath{\langle#1\rangle} }

\shortauthors{ENOTO ET AL.}
\shorttitle{
{\it Suzaku} observation of a symbiotic X-ray binary 4U~1954+319 
}

\begin{document}
\title{
Spectral and Timing Nature of the Symbiotic X-ray Binary 4U~1954+319: \\
The Slowest Rotating Neutron Star in an X-ray Binary System
}

\author{
Teruaki Enoto\altaffilmark{1,2}
Makoto Sasano\altaffilmark{3},
Shin'ya Yamada\altaffilmark{2},  
Toru Tamagawa\altaffilmark{2},  
Kazuo Makishima\altaffilmark{2,3}, 
\\
Katja Pottschmidt\altaffilmark{4,5}, 
Diana Marcu\altaffilmark{4,5}, 
Robin H.D. Corbet\altaffilmark{1, 5}, 
Felix Fuerst\altaffilmark{6}, 
and 
J\"{o}rn Wilms\altaffilmark{7}
}
\altaffiltext{1}{
NASA Goddard Space Flight Center, Astrophysics Science Division, Code 662, 
        Greenbelt, MD 20771, USA;
        teruaki.enoto@nasa.gov}
\altaffiltext{2}{
High Energy Astrophysics Laboratory,
RIKEN Nishina Center, 2-1 Hirosawa, Wako, Saitama 351-0198, Japan
    }
\altaffiltext{3}{Department of Physics, University of Tokyo,
    7-3-1 Hongo, Bunkyo-ku, Tokyo, 113-0033, Japan}
\altaffiltext{4}{
NASA Goddard Space Flight Center, Astrophysics Science Division, 
Code 661, Greenbelt, MD 20771, USA
}
\altaffiltext{5}{
CRESST \& University of Maryland Baltimore County, Baltimore, MD 21250, USA
}
\altaffiltext{6}{
Cahill Center for Astronomy and Astrophysics, California Institute of Technology, Pasadena, CA 91125
}
\altaffiltext{7}{
Dr.\ Remeis-Sternwarte and Erlangen Centre for Astroparticle Physics,
Universit\"at Erlangen-N\"urnberg, Sternwartstr.~7, 96049 Bamberg
}

\begin{abstract}
  The symbiotic X-ray binary 4U~1954+319 
  	is a rare system hosting a peculiar neutron star (NS) and 
	an M-type optical companion. 
Its $\sim$5.4\,h NS spin period is 
	the longest among all known accretion-powered pulsars 
	and exhibited large ($\sim$7\%) fluctuations over 8 years. 
A spin trend transition
	was detected with \textit{Swift}/BAT 
	around an X-ray brightening in 2012. 	
The source was in 	
	quiescent and bright states 
	before and after this outburst
	based on 60 ks \textit{Suzaku} observations 
	in 2011 and 2012.
The observed continuum is well described by 
	a Comptonized model
	with the addition of a narrow 6.4 keV Fe K$\alpha$ line during the outburst.
Spectral similarities to slowly rotating pulsars in high-mass X-ray binaries,
	its high pulsed fraction ($\sim$60--80\%), 
	and the location in the Corbet diagram
	favor 
	high $B$-field ($\gtrsim$$10^{12}$ G)
	over a weak field as in low-mass X-ray binaries.
The observed low X-ray luminosity ($10^{33}$--$10^{35}\,\mathrm{erg}\,\mathrm{s}^{-1}$),
	probable wide orbit, and a slow stellar wind
	of this SyXB
	make quasi-spherical accretion  
	in the subsonic settling regime a plausible model.
Assuming a $\sim$$10^{13}$\,G NS, this scheme 
	can explain the $\sim$5.4\,h equilibrium rotation 
	without employing the magnetar-like field ($\sim$$10^{16}$\,G) 
	required in the disk accretion case.
The time-scales of multiple irregular flares ($\sim$$50$\,s) 
	can also be attributed to 
	the free-fall time 
	from the Alfv\'en shell for a $\sim$$10^{13}$\,G field.
A physical interpretation of  SyXBs 
	beyond the canonical binary classifications is discussed.
\end{abstract}


\section{INTRODUCTION}
\label{Introduction}

Symbiotic X-ray binaries (SyXBs) are a growing new subclass of X-ray
binaries which consist of an X-ray bright neutron star (NS) and an
M-type giant primary star. 
Originally SyXBs were seen as a subclass of low-mass X-ray
binaries (LMXB), where accretion typically is from a K-type optical
counterpart with a mass $M_\mathrm{c}\lesssim 1\,M_\odot$. 
It took more than 30 years, however, to recognize
that their properties are different from LMXB and also from
those of high-mass binaries (HXMB), i.e., systems where the mass donor
is an O- or B-type giant with $M_\mathrm{c}\gtrsim 10\,M_\odot$.
SyXBs are now considered as a separate
subclass of X-ray binaries. They are named in analogy to symbiotic
binaries, which consist of a white dwarf and an M giant companion, but
their nature has not yet been understood.
        
\begin{figure*}
\includegraphics[width=80mm]{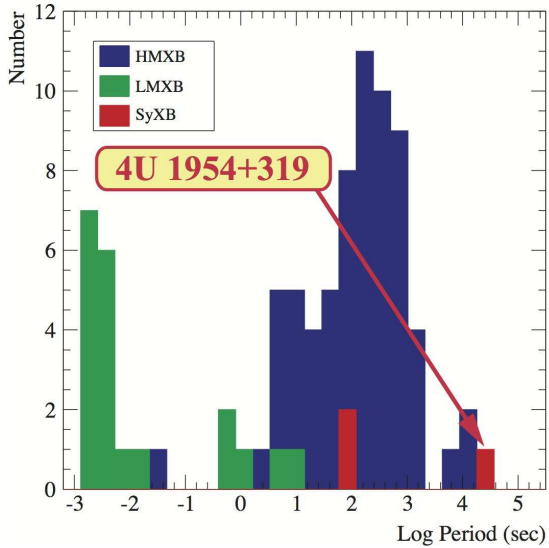}
\caption{ Distribution of NS rotation periods in HMXBs
  \citep{Liu2006A&A}, LMXBs \citep{Liu2007A&A}, and confirmed SyXBs.
  Measured rotation periods of SyXBs are      
  $\sim$18300 s (4U~1954+319),
  $\sim$110 s (Sct~X-1),
  and 
  $\sim$150 s (GX~1+4).  
    The 4U~1954+319 is in the HMXB range.}
\label{fig:period_distribution}
\end{figure*}

Some SyXBs were known 
	as famous X-ray sources 
	since the early X-ray observations: 
	GX~1+4 \citep{Davidsen1977ApJ, Makishima1988Nature,  Chakrabarty1997ApJ}, 
	4U~1700+24 \citep{Garcia1983ApJ,Masetti2002A&A}, 
	and Sct~X-1 \citep{Kaplan2007ApJ}.
With the discovery of  new sources such as IGR~J16194$-$2810,
	there are now nearly 5 SyXBs and some candidates in our Galaxy; 
	e.g., 5 objects out of 10 candidates (Table~1 in \citealt{Lu2012MNRAS})
	have been confirmed as SyXBs, while three were ruled out 
	(see e.g., \citealt{Masetti2012A&A...544A.114M}).
Observational features of the SyXB class are 
i) quite long orbital periods\footnote{For
	many SyXB, orbital period has not yet been known 
	nor confirmed in long-term X-ray studies (e.g., \citealt{Corbet2008ApJ}).},
	e.g., 1161\,d in GX~1+4. 
ii) long NS spin periods over
	$\sim$110--18300\,s as shown in Fig.~\ref{fig:period_distribution},
	and 
iii) high X-ray variability 
	ranging from short to long timescales ($\sim$1\,s to up to a year).

SyXB X-ray spectra are in general well described by an absorbed
power-law with a photon index $\Gamma$$\sim$0.5--2.0 and a high energy
roll-over, which is often modified in the soft X-rays by strong absorption. 
This spectrum resembles that of X-ray pulsars in HMXBs, 
and thus, SyXBs have been interpreted in the literature as NSs
with $\sim 10^{12}$ G field in HMXBs. 
As another interpretation,
following the classical LMXB classification based on optical companions, 
SyXBs were also interpreted as NSs in the low luminosity low/hard
state of LMXB as having weak magnetic field ($B$-field), e.g., $\lesssim 10^{10}$\,G
\citep{Nagae2008,Kitamura2013PASJ}. 
In any of these cases, the NSs in
SyXB are expected to have $B$-fields strengths of around
$10^{12}$\,G or much lower (e.g., $\lesssim 10^{10}$\,G).

The canonical $B$-fields implied by the spectral analogy with
HMXB and LMXB, however, are in contradiction to the fields implied by
their timing behavior. The slow NS pulsations in SyXBs are sometimes
in spin equilibrium or show a large spin-up rate
\citep{Makishima1988Nature, Gonzalez2012A&A...537A..66G}. If
interpreted in terms of standard disk accretion torque theory
\citep{Ghosh1979ApJ}, the slow rotation and torque reversals imply
extraordinary strong dipole fields ($\sim$$10^{14-16}$\,G), which are
close to or exceed the quantum critical field,
$B_\mathrm{cr}=m_\mathrm{e}^2c^3/(e\hbar)=4.4\times 10^{13}$\,G. 
Such a magnetar scenario for binary systems has been proposed, e.g., to
explain the 1.6\,h pulsation and the high spin-up rate of the SyXB
IGR~J16358$-$4726 \citep{Patel2007ApJ}. Similar scenarios have also been 
discussed for slowly rotating X-ray pulsars in HMXBs such as 4U~0114+65
\citep{Li1999ApJ...513L..45L}  or 4U~2206+54 \citep{Reig2012MNRAS}, 
highly variable supergiant fast X-ray transients (SFXTs;
\citealt{Bozzo2008ApJ}), and some slowly rotating NSs 
in the Small Magellanic Cloud \citep{2014MNRAS.437.3664H, Klus2014MNRAS.437.3863K}.
Except for the timing properties, however, no
direct evidence for ultra-strong $B$-fields has been found so
far.

As shown by \citet{Shakura2012MNRAS} and \citet{2012arXiv1212.2841P},
the extremely strong fields can be avoided when one assumes that the
accretion does not happen through an accretion disk but via
quasi-spherical accretion. In these models matter is gravitationally
captured, e.g., from the donor's stellar wind. For low mass accretion
rates, $\dot{M}\lesssim 4\times 10^{16}\,\mathrm{g}\,\mathrm{s}^{-1}$,
gravitationally captured matter subsonically accretes onto the NS
forming an extended, quasi-spherical shell around the magnetosphere,
where large scale convective motions and subsonic turbulence lead to
accretion onto the NS. Contrary to the canonical high luminosity
Bondi-Hoyle accretion, where matter supersonically accretes and the
sign of angular momentum (prograde or retrograde) determines the spin
behavior \citep{Perna2006ApJ}, for quasi-spherical accretion the
spin-up/spin-down is determined by the specific angular momentum of
matter at the magnetospheric boundary and the angular velocity of the
NS.

In this paper we discuss \textit{Suzaku} observations of the SyXB
4U~1954+319 hosting the slowest rotating NS in an X-ray binary system. 
This source was originally discovered by \textit{Uhuru}
\citep{Forman1978ApJS}, and observed in the 1980s with \textit{Ariel}
\citep{Warwick1981MNRAS}, \textit{EXOSAT}, and \textit{Ginga}
\citep{Tweedy1989ESASP}. These observations suggested that the source
was a wind accreting HMXB with a very inhomogeneous wind
\citep{Tweedy1989ESASP}, although no optical counterpart was known at
that time. The source had virtually been forgotten until, after twenty
years of no observations, \textit{Chandra}'s localization of the
source position allowed the optical identification of the
donor, which was found to be an M4-5 III star ($M_{\rm c}\sim
1.2M_{\odot}$) at a distance of 1.7\,kpc \citep{Masetti2006A&A}.
Therefore 4U~1954+319 was re-classified as a SyXB.

Further progress in our understanding of the characteristics came with
the discovery of a $\sim$5\,h periodicity in \textit{Swift}/BAT
monitoring data by \citet{Corbet2006ATel, Corbet2008ApJ}. This period is strongly
variable \citep{Marcu2011ApJ}. So far, no orbital period has been
discovered, although \citet{Mattana2006A&A} argue that the lower limit
of the orbital period is $\sim$400\,d. These parameters are
incompatible with typical white dwarf systems, such that the compact
object in 4U~1954+319 is likely a NS. As shown in
Figure~\ref{fig:period_distribution}, the extremely long spin period
makes 4U~1954+319 the slowest rotating accretion-powered NS
system known to date\footnote{A long periodicity, $P\sim 6.67$ hour,
  was also detected from the central object 1E161348$-$5055 in the
  young shell-type supernova remnant RCW 103. This object, however,
  has not been confirmed as an accretion-powered pulsar.}, and a
prominent example to understand the SyXB class.

Currently, 
	no consensus has yet been obtained on the accretion models 
	for SyXBs, or the $B$-field of their NSs.
Revisiting and
developing the quasi-spherical model further requires the spectral and
timing investigations of SyXBs. In this paper, we report  on the broad-band
timing and spectral properties of 4U~1954+319 and discuss a possible accretion
model beyond the conventional HMXB and LMXB classification.
        
\section{OBSERVATION AND DATA REDUCTION}\label{sec:obs}
\label{observation and data processing}

\subsection{Suzaku observations}

There are two pointed \textit{Suzaku} \citep{Mitsuda2007PASJ}
observations of 4U~1954+319.  
The first one, in AO6, was
conducted on 2011 October 23--24 for a gross on-source time of 117\,ks.
Nearly one year later, on 2012
October~5, MAXI monitoring detected a brightening of this
source\footnote{Trigger ID 6206800002:
  \url{http://maxi.riken.jp/pipermail/x-ray-star/2012-October/000187.html}},
with the X-ray intensity reaching $\sim$15 mCrab in the
3--10\,keV band. About 27\,days later, on 2012 November 1/2, a
follow-up ToO observation was performed with \textit{Suzaku} for a
gross on-source time of 120\,ks. 
Table~\ref{tab:Obslog} summarizes the two \textit{Suzaku} observations.

The X-ray Imaging Spectrometer \citep[XIS;][]{Koyama2007PASJ} on board
\textit{Suzaku}, coupled to the X-ray Telescopes
\citep[XRT;][]{Serlemitsos2007PASJ}, consists of two
front-illuminated (FI) CCD detectors, XIS0 and XIS3, and one
back-illuminated sensor (XIS1). The CCDs are sensitive in the
0.2--12\,keV range and, when operated in the full window mode, read
out every 8\,s. We also used data from the Hard X-ray Detector
\citep[HXD][]{Takahashi2007PASJ}, which consists of Silicon PIN diodes
(HXD-PIN) and GSO scintillators (HXD-GSO), and cover the 10--70\,keV
and 50--600\,keV bands, respectively. The HXD was operated in the
nominal mode during both observations with 61\,$\mu$s time resolution.
The bright X-ray source Cygnus X-1, which is only $3\fdg15$ from
4U~1954+319, is still outside the HXD-PIN's tightly collimated
field of view (FOV, $34'\times34'$ FWHM), but within the FOV of the
HXD-GSO ($4^\circ\times4^\circ$). Thus, we did not use the HXD-GSO
data in the following analyses due to a possible contamination from
Cyg X-1. Based on the INTEGRAL source catalog, there is no other
strong contamination source within the HXD-PIN FoV.

\subsection{Data processing}
\label{data_processing}

We analyzed the XIS and HXD datasets as obtained from the standard
pipeline processing, revision 2.7 and 2.8 for the 2011 and 2012 data
sets, respectively. All data analyses were performed using HEADAS
version 6.11 or later. Following the standard \textit{Suzaku}
screening criteria, X-ray events were discarded if one of the
following conditions held true: (a) they were recorded during the South
Atlantic Anomaly, $\texttt{TSAA}\le 436\,\mathrm{s}$ and
$\texttt{TSAA} \le 500\,\mathrm{s}$ for the XIS and the HXD,
respectively, and later than 180\,s before the next passage for the
HXD,
(b) while the satellite was in regions of low geomagnetic cutoff
rigidity ($\le$6\,GV), (c) while the target was closer than $5^\circ$
to the Earth's rim
(d) if the instantaneous pointing direction deviated more than $1.5'$
from the mean, or (e) during periods of telemetry saturation. After
screening, the net exposures for the 2011 observation were 60.2\,ks
(XIS) and 53.7\,ks (HXD), while those in 2012 were 60.8\,ks (XIS) and
50.5\,ks (HXD), respectively.

On-source XIS events were extracted from a region of
$2\farcm5$ radius centered on the source. Background events were
derived from an annulus $4\farcm0$--$7\farcm5$ around the source. From
the HXD-PIN data, we subtracted the Non X-ray Background (NXB) created
with the LCFITD method \citep{Fukazawa2009PASJ} and filtered following
the same screening criteria as those used for the on-source data. The
Cosmic X-ray Background (CXB) was modeled as described by
\citet{Enoto2010PASJ} with the same spectral shape as given by
\citet{Moretti2009A&A}. In the following analyses, unless otherwise
specified, all uncertainties are given at the 68\% ($1\sigma$)
confidence level.

\subsection{Reduction of the X-ray spectra}\label{sec:analysis}

For the spectral analysis we summed the two XIS-FI sensors and utilized
XIS data from the 0.8--10\,keV band and the HXD-PIN data from the
15--70\,keV band. The 1.7--1.9\,keV XIS data are discarded due to
calibration uncertainties. Uncertainties in the flux normalization of
the instruments are taken into account with a multiplicative constant.
The cross-normalization factor of HXD-PIN relative to XIS-FI was fixed at 
1.171 based on the XIS-nominal Crab Nebula calibration
\citep{Maeda2008SuzakuMemo}, while that of XIS-FI and XIS-BI was allowed to
differ by up to 1\%. A 0.5\% systematic uncertainty was added in
quadrature to the 2012 XIS spectral bins in order to absorb
instrumental calibration uncertainties. The response matrix
files (RMFs) and auxiliary response file (ARFs) for the XIS were
produced using the FTOOLs \texttt{xisrmfgen} and \texttt{xissimarfgen}
\citep{Ishisaki2007PASJ}, while standard detector response files
(epoch 11) were used for the HXD-PIN.

\section{Suzaku Observations of 4U~1954+319}
\subsection{Long-term variability}
\label{Long-term variability}

\begin{figure*}
\includegraphics[width=183mm]{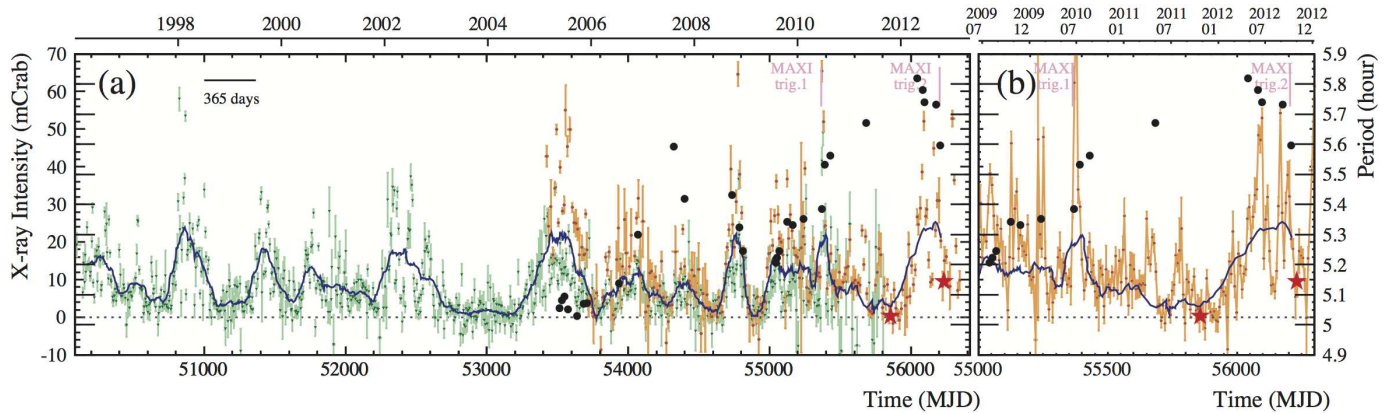}
\caption{
 Long-term light curve of 4U~1954+319. 
 (a) 2--10 keV \textit{RXTE}/ASM (triangle, green) and 15--50
  keV \textit{Swift}/BAT (filled circle, orange) light curves of
  4U~1954+319 since 1994. The solid blue curve represents a running
  average of the X-ray intensity with the window size of 20\,d. The
  detected pulse period history by the \textit{Swift}/BAT is overlaid
  as black filled circles (right hand axis). Two MAXI detections of
  the flaring activity of this source are also shown: MAXI trig 1 on
  2012 June 29 \citep{Sugizaki2010ATel} and trig 2 on 2012 October 5
  (see text). 
The two {\it Suzaku} observations are indicated as red star symbols.
  (b) Enlarged version of the light curves around the two
  \textit{Suzaku} observations, showing the 7\,days averaged X-ray
  intensity from \textit{Swift}/BAT (orange). }
\label{fig:long-term_lc}
\end{figure*}

Figure~\ref{fig:long-term_lc} shows 
	the long-term variability of the X-ray intensity of 4U~1954+319, 
	covering $\sim$17 years since 1994.
X-ray count rates
	were re-binned to 6 and 12\,days resolution
	and converted into the mCrab unit using conversion factors 
	provided by the \textit{RXTE}/ASM and \textit{Swift}/BAT teams\footnote{The 
	long-term \textit{RXTE}/ASM monitoring data
	  are provided by the \textit{RXTE}/ASM team
	  (\url{http://xte.mit.edu}), while the \textit{Swift}/BAT transient
	  monitor results provided by the Swift/BAT team
	  (\url{http://heasarc.gsfc.nasa.gov/docs/swift/results/transients/}).
	The reported conversion factors, assuming a Crab-like spectrum, are 
	$7.5\times 10^{-2}\,\mathrm{counts}\,\mathrm{s}^{-1}\,\mathrm{mCrab}^{-1}$
	and 
	$2.2\times 10^{-4}\,\mathrm{counts}\,\mathrm{cm}^{-2}\,\mathrm{s}^{-1}\,\mathrm{mCrab}^{-1}$, 
	for ASM and BAT, respectively.}.
As shown in the 20\,d running average of the fluxes
(Fig.~\ref{fig:long-term_lc}, blue curve), the X-ray intensity of
4U~1954+319 is highly variable on a time scale of years, with fluxes
changing from a few mCrab to at least 20 mCrab or more. In addition,
faster variability on timescales of weeks or days is also seen with irregular flares.
        
Figure~\ref{fig:long-term_lc} also shows the spin history of
4U~1954+319 as measured with \textit{Swift}/BAT. 
We determined the pulse periods using standard epoch folding \citep{leahy:83b} and using
pulse profiles with 19\,phase bins.
Spin values are shown with filled circles if pulses are detected with
$\chi^2_\mathrm{red}>3$. The measured period history is consistent
with previous work covering data taken before 2010
\citep{Marcu2011ApJ}. The present work extends the analysis until the
end of 2012 and shows the detection of a new change from spin-down to
spin-up which occurred around 2012 March (MJD$\sim$56000), when the
average X-ray intensity increased to $\sim$25\,mCrab. The analysis
shows that the first and second \textit{Suzaku} observations were
taken during the spin-down and spin-up phases, respectively. The former 
observation was almost in quiescence ($\sim$0.3\,mCrab), before the
outburst, while the latter was during a brighter phase after the
flaring ($\sim$10\,mCrab) and the spin-transition.

\subsection{Timing analysis}\label{sec:timing}
\label{Light curves and pulse profiles}

\begin{figure*}
\includegraphics[width=183mm]{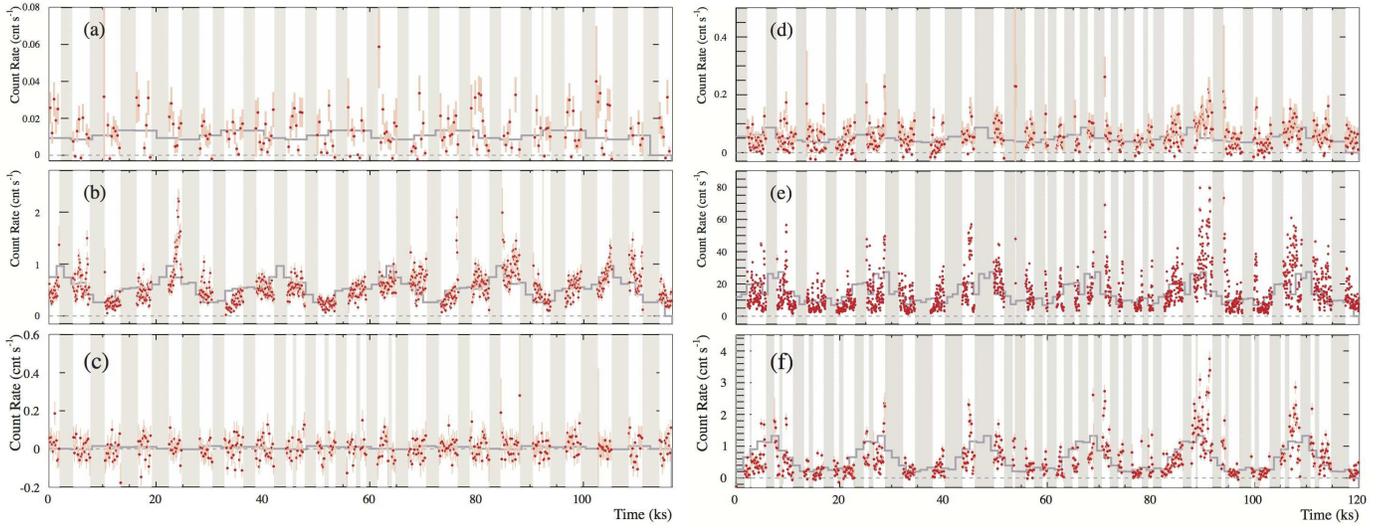}
\caption{ Background-subtracted light curves of 4U~1954+319 obtained
  with \textit{Suzaku} in 2011 October (left) and 2012 November
  (right). Gray strips represent the non good time intervals (non-GTIs)
  during the observation. From top to bottom, the panels show the
  0.5--1.0\,keV XIS (binning 360\,s [2011] and 120\,s [2012]), 1--10
  keV XIS (64\,s and 32\,s), and 15--70\,keV HXD-PIN (160\,s and 80\,s) count
  rates. The XIS background was subtracted as described in
  \S\ref{data_processing}, while the simulated NXB
  ($\sim$$0.28\,\mathrm{count}\,\mathrm{s}^{-1}$) and CXB
  ($\sim$$0.02\,\mathrm{counts}\,\mathrm{s}^{-1}$) were subtracted
  from the dead-time corrected HXD-PIN data. Estimated pulse shapes
  are overlaid in each panel, using the \textit{Swift}/BAT period
  described in \S\ref{Light curves and pulse profiles}. }
\label{fig:suzaku_lc}
\end{figure*}

\begin{figure*}
\includegraphics[width=80mm]{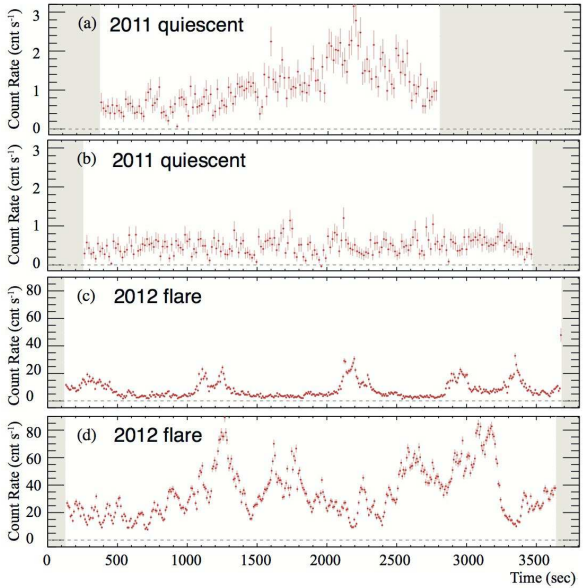}
\caption{ XIS 1--10\,keV light curves during the 2011 (a,
  b) and the 2012 observation (c, d). Data are binned to 8\,s resolution.
 Time durations are 
 (a) 22.0--25.8 ks,
 (b) 55.8--59.6 ks,
 (c) 13.6--17.4 ks,
 and (d) 88.2--92.0 ks in Fig.~\ref{fig:suzaku_lc}.
  }
\label{fig:suzaku_lc_enlarge}
\end{figure*}

\begin{figure*}
\includegraphics[width=100mm]{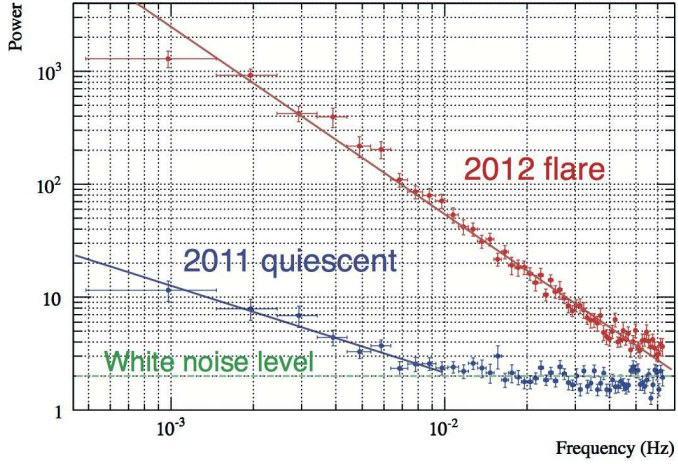}
\caption{ Power spectrum densities (PSD) of the 0.8--10 keV XIS data in 2011
  and 2012 observations as calculated with the \textit{FTOOL}
  \texttt{powspec} using the background-subtracted 8\,s binned light curve. 
 The PSDs are normalized such that the white noise level corresponds to 2. 
  The best-fit power-law models are shown with slopes of
  $-0.77$ and $-1.66$ in 2011 and 2012, respectively,
  when fitted above the white noise level.  
  }
\label{fig:power_spectrum}
\end{figure*}

Figure~\ref{fig:suzaku_lc} shows the XIS and HXD-PIN light curves of
4U~1954+319 from the two observations. 
Similar to the behavior observed on long timescales
(Fig.~\ref{fig:long-term_lc}), the X-ray light curve is highly
variable also on short time scales. 
Figure~\ref{fig:suzaku_lc_enlarge}
	further illustrates the enlarged light curve.
The X-ray fluctuation was moderate in 2011, while spiky
flare events were seen in 2012 with $\lesssim$100 s scale, showing a
rapid flux change by an order of magnitude or more.
During the 2012 outburst (Fig.~\ref{fig:suzaku_lc_enlarge}),
	the spiky flares become enhanced in the latter part of observation. 
An increase of the short-term
fluctuation is also seen in the power spectral density (PSD)
in Figure~\ref{fig:power_spectrum}. 
The PSD continuum can be described by
a power law components with best fit slopes of $-0.77$ in 2011 and
$-1.66$ in 2012. There are no signatures of clear short-term periods
or quasi periodic oscillations. 
No low-frequency roll over was detected, either.

Epoch folding of the 2012 XIS and HXD data yielded a spin period of
$5.68\pm0.28$\,h. However, the period determination
with \textit{Suzaku} is not accurate enough, due to the limited exposure which
corresponds to only $\sim$6 cycles of the NS rotation. Thus, we
searched the \textit{Swift}/BAT monitoring data (\S\ref{Long-term
  variability}) for the spin period using intervals of $\pm60$\,d and
$\pm400$\,d around the \textit{Suzaku} pointings. 
The pulse periods derived from these data are $P=5.70\pm0.01$\,h
($\chi^2/\nu=104.7/14=7.5$) and $P=5.76\pm0.01$\,h
($\chi^2/\nu=72.3/14=5.2$), in 2011 and 2012, respectively, where 15
phase bins were used for the epoch folding. 
Average pulse profiles, 
	obtained at these periods, 
	are superposed in Fig.~\ref{fig:suzaku_lc}
	on the {\it Suzaku} light curves.
Even though sporadic flares are rather strong,
	we can trace individual pulses in most cases.	

\begin{figure*}
\includegraphics[width=86mm]{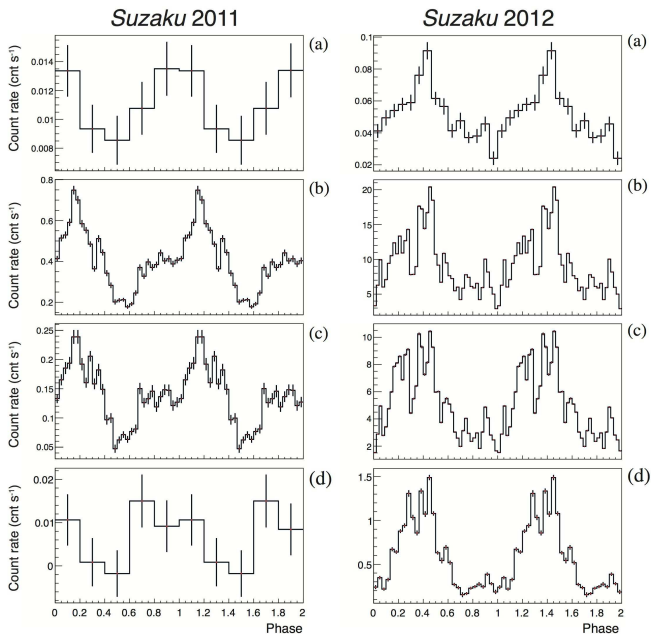}
\caption{ Pulse profiles of 4U~1954+319 folded with 5.70\,h and 5.76\,h
  in 2011 (left) and 2012 (right), respectively. Panels a, b, c, and d
  represent the 0.5--1.0\,keV XIS, the 1--5\,keV XIS, the 5--10\,keV
  XIS, and the 15--70\,keV HXD-PIN energy bands, respectively.
  Backgrounds were subtracted in the same way as for
  Fig.~\ref{fig:suzaku_lc}. }
\label{fig:folding}
\end{figure*}

\begin{figure*}
\includegraphics[height=7cm]{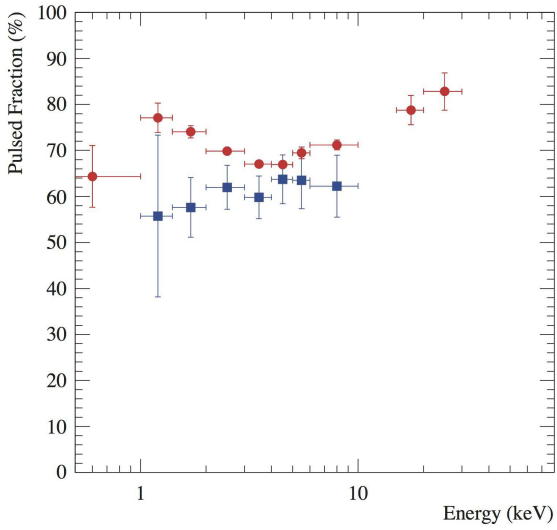}
\caption{ Pulsed fraction of 4U~1954+319 in 2011 (blue) and 2012
  (red). }
\label{fig:pulse_profile}
\end{figure*}

Figure~\ref{fig:folding} shows the pulse profiles of the two {\it Suzaku} observations. 
The pulse profile in 2012 is approximately sinusoidal below 1\,keV, 
with a single peak at phase $\sim$0.4. A second peak at phase $\sim$0.2 appears at higher energies.
Figure~\ref{fig:pulse_profile} shows the energy dependence of 
	the pulsed fraction of 4U~1954+319, defined as
$\mathrm{PF}=(R_\mathrm{max}-R_\mathrm{min})/(R_\mathrm{max}+R_\mathrm{min})$,
where $R_\mathrm{max}$ and $R_\mathrm{min}$ are the maximum and
minimum of the pulse profile. The fraction is 60--80\% in
the 0.5--30\,keV band, and increases with energy. This is a typical
trend often seen in X-ray pulsars in HMXBs
\citep{Lutovinov2008AIPC.1054..191L}.

\subsection{X-ray spectrum during the flare in 2012}
\begin{figure*}
\includegraphics[scale=0.4]{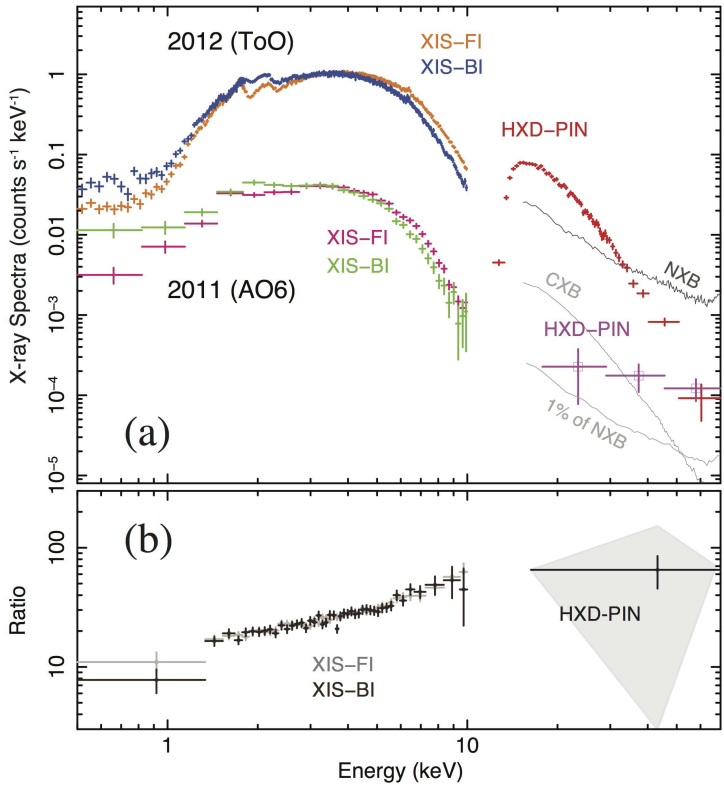}
\caption{ (a) Time-averaged, background subtracted raw XIS-FI, XIS-BI,
  and HXD-PIN spectra of 4U~1954+319 during the two \textit{Suzaku}
  pointings in count rate space. The modeled NXB, CXB, and 1\% of NXB
  (a typical uncertainty of the NXB modeling) are also indicated. (b)
  Spectral ratio of the 2012 ToO data to those from the 2011
  observation. The gray diamond of the HXD-PIN point represents an
  error region including 0.3\% systematic uncertainties. }
\label{fig:raw_spec}
\end{figure*}

Figure~\ref{fig:raw_spec}a shows \textit{Suzaku}
spectra of 4U~1954+319 obtained on the two occasions. 
We superpose the modeled HXD-PIN NXB and CXB, 
together with the typical 1\% systematic uncertainty of the NXB. The
HXD-PIN detection is marginal in 2011. 
The ratio between the two spectra is shown in
Fig.~\ref{fig:raw_spec}b. In 2012, 4U~1954+319 was brighter by nearly
1--2 orders of magnitude,
and exhibited harder spectra.

\begin{figure*}
\includegraphics[scale=0.32]{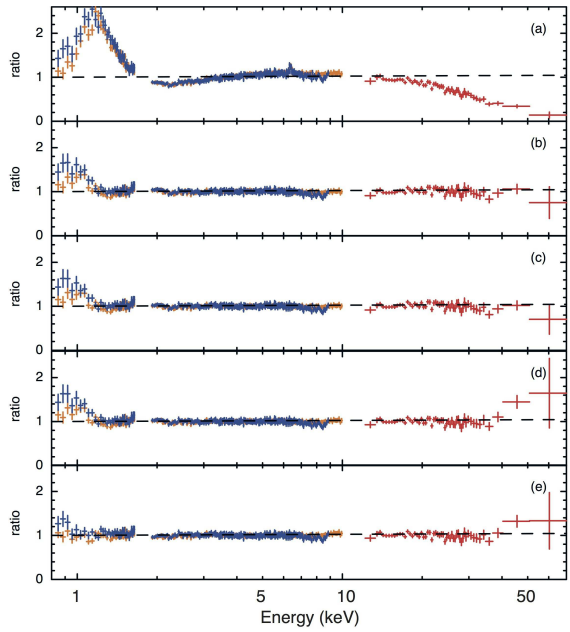}
\caption{ Ratios of the best fit spectral model to the 2012
  \textit{Suzaku} data, using (a) a single power-law, (b) a cutoff
  power-law, (c) a high energy cutoff, (d) the NPEX, 
  and (e) the \texttt{comptt}.
  The 6.4 keV gaussian emission is included in panels b, c, d, and e,
  while a soft blackbody component ($kT\sim0.13$\,keV) is added in panels b, c, and d.
}
\label{fig:spectral_fit_ratio}
\end{figure*}

We now turn to modeling the X-ray spectrum in Figure~\ref{fig:spectral_fit_ratio}, 
and summarize the best fit parameters 
in Table~\ref{tab:TimeAveragedSpectra2012}. 
Fitting the data with a single, absorbed 
power law leaves a soft excess below 2\,keV, an iron line at $\sim$6.4\,keV,
and a spectral roll over (Fig.~\ref{fig:spectral_fit_ratio}a). 
Here we use the latest interstellar absorption model \texttt{tbnew}
\citep{Wilms2000ApJ...542..914W} 
with an improved abundance \texttt{wilm} and cross section 
tables \texttt{vern} \citep{Verner1996ApJ...465..487V}.
Due to higher data quality compared to earlier {\it INTEGRAL}
measurements, the broken power-law model used by
\citet{Furst2011arXiv} and \citet{Marcu2011ApJ} was not successful 
($\chi^2_{\nu}=2373.6/697=3.40$), either. 
The exponentially cutoff power-law with an
additional soft blackbody and an Fe-line, improved the fit to
$\chi^2/\nu=971.5/696=1.40$ (Fig.~\ref{fig:spectral_fit_ratio}b) with a photon
index of $\Gamma=1.21$, a cutoff energy $E_\mathrm{cut}=20$\,keV, and
a blackbody temperature of $kT_\mathrm{bb}=0.13$\,keV. The spectral
parameters are similar to those found by \citet{Mattana2006A&A}
by applying the same model to the {\it BeppoSAX} data, 
although our fit is not satisfactory.

As alternative representations of the roll-over of the continuum, we
employed two empirical models: 1) a high energy cutoff model
of the form (in \texttt{XSPEC} notation)
\begin{equation}
\texttt{phabs}\times(\texttt{bbodyrad}+\texttt{gauss}+\texttt{powerlaw}\times
\texttt{highecut})
\end{equation}
where
\begin{equation}
\texttt{highecut} = 
\begin{cases}
1 & \text{for $E<E_\mathrm{c}$} \\
\exp [ {(E_\mathrm{c}-E)}/{E_\mathrm{f}} ] & \text{for $E\ge
  E_\mathrm{c}$}
\end{cases}
\end{equation}
with the cutoff energy $E_\mathrm{c}\sim4.2$\,keV and e-folding energy 
$E_\mathrm{f}\sim21.0$\,keV. We also modeled the data using the
2) a Negative and Positive power law with an EXponential (NPEX),
\begin{equation}
\texttt{phabs}\times(\texttt{bbodyrad}+\texttt{gauss}+\texttt{cutoffpl}+\texttt{cutoffpl})
\end{equation} 
	which is often utilized 
	to represent broad-band continuum from the accretion
	column of highly magnetized NSs \citep{Mihara1995PhDT,Makishima1999ApJ...525..978M,Enoto2008PASJ}.
The cutoff energy and the photon index of the NPEX continuum 
	were obtained as 6.6\,keV and 0.77, respectively,
	and the blackbody temperature as 0.14 keV.
As shown in Fig.~\ref{fig:spectral_fit_ratio}c, d, and
Table~\ref{tab:TimeAveragedSpectra2012}, both models reproduced the
broadband spectra quite well ($\chi^2_\nu\sim 1.2$). 
If we exclude the soft blackbody, $\chi^2$ becomes worse
($\chi^2_{\nu}>3.5$). Replacing the black body component with a
diffuse plasma emission model (\texttt{apec}\footnote{Astrophysical Plasma
  Emission Code (APEC), a collisonally-ionzied diffuse gas model
  \url{http://atomdb.org/}}; \citealt{Smith2001ApJ...556L..91S}), 
did not give acceptable fits ($\chi^2_{\nu}>1.9$).
        
\begin{figure*}
\includegraphics[scale=0.45]{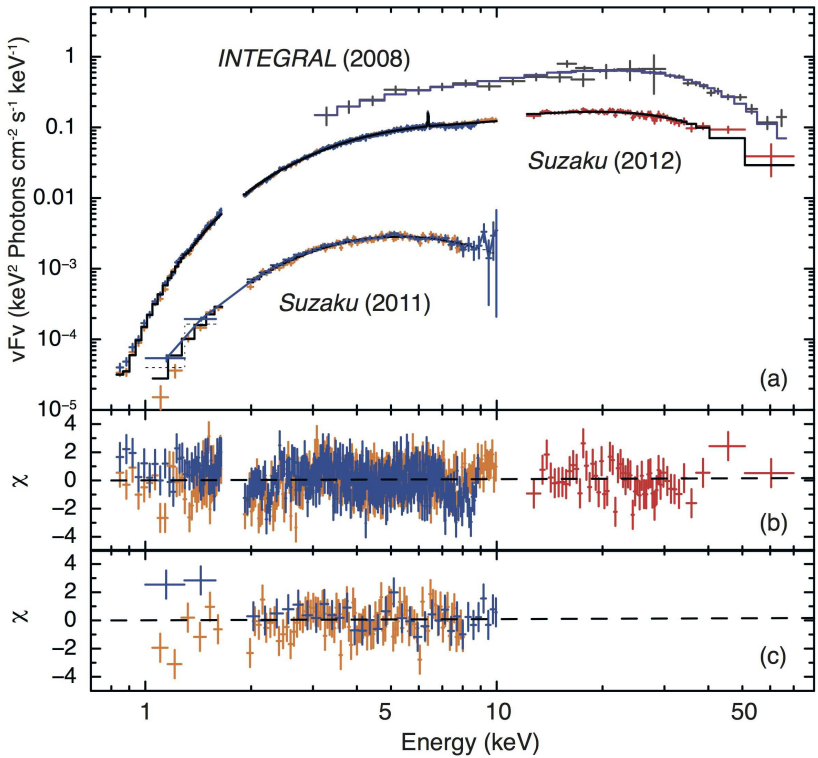}
\caption{ (a) X-ray spectra of 4U~1954+319 in $\nu F_{\nu}$ form
  observed during the 2011 quiescent and 2012 flare-up, utilizing the
  \texttt{comptt} continuum (Model C). 
  For comparison,
  {\it INTEGRAL} spectra, JEM-X (3-30 keV) and 
  ISGRI (20-80 keV) for the flaring episode in 2008, 
  are overlaid with a \texttt{comptt} model fit as used by \cite{Marcu2011ApJ},
  (b) residuals of the
  \texttt{comptt} model in 2012. (c) the same as panel b but for the
  2011 data. }
\label{fig:nuFnu}
\end{figure*}

Finally, as a more physically-based interpretation, we also tried an
optically thick Comptonizaiton model, \texttt{compTT} 
(\citealt{Titarchuk1995ApJ...450..876T}). 
This gave an acceptable fit,
$\chi^2_{\nu}=780.8/697=1.12$, 
with a null hypothesis probability of 0.02
(Fig.~\ref{fig:spectral_fit_ratio}e). 
We therefore adopt this model as
	a canonical continuum model of 4U~1954+319.
The derived best-fit parameters
	are a soft photon (Wien) temperature of $kT_0\sim 1$\,keV, 
	plasma temperature $kT_\mathrm{e}\sim 8$\,keV, 
	and an optical depth of $\tau \sim 9.0$. 
These derived $kT_0$ and $kT_\mathrm{e}$ values are
	within the typical range of the \texttt{comptt} model of this source,
	$kT_0\sim 0.6$--$1.3$\,keV and $kT_\mathrm{e}\sim 3.0$--$13$\,keV,
	 by the multi X-ray satellite studies  \citep{Masetti2007A&A...464..277M} 
	 and 
	 by the {\it INTEGRAL} observation \citep{Marcu2011ApJ}.
Combining the previous two successful models,
	the broad-band spectrum is explained by the Comptonized model.   
Figure~\ref{fig:nuFnu} gives a $\nu F_{\nu}$ representation of the data and the model.	
The 2--10\,keV and 1--70\,keV absorbed X-ray
fluxes are $(1.90\pm 0.01) \times 10^{-10}$ and $(6.0\pm 0.1) \times
10^{-10}\,\mathrm{erg}\,\mathrm{s}^{-1}\,\mathrm{cm}^{-2}$,
translating to absorption-corrected X-ray fluxes of
$(2.03\pm0.01)\times 10^{-10}$ and 
$(6.3\pm0.1)\times 10^{-10}\,\mathrm{erg}\,\mathrm{s}^{-1}\,\mathrm{cm}^{-2}$,
respectively.


\subsection{Iron line and search for CRSF } 

The 6.4 keV neutral Fe K$\alpha$ line emission is obvious in
Fig.~\ref{fig:spectral_fit_ratio}a. In the \texttt{compTT} continuum
fits, its center energy was found at $6.375^{+0.010}_{-0.008}$ keV
with an equivalent width of $\mathrm{EW}=28.6\pm 0.3$\,eV and a total
line flux of $I_\mathrm{Fe}=7.5^{+0.8}_{-0.7} \times
10^{-5}\,\mathrm{ph}\,\mathrm{cm}^{-2}\,\mathrm{s}^{-1}$. 
The line is
narrow with a width $\sigma < 50$\,eV ($1\sigma$), corresponding to a
Doppler width $v_{\infty}\lesssim 2400\,\mathrm{km}\,\mathrm{s}^{-1}$.

Detailed inspection of the spectrum indicates a shallow dent at $\sim$7.1 keV,
	and if we let the iron abundance free,
	the data favor the 2.4-times larger iron abundance relative to the default 
	improving the fit  by $\triangle \chi^2=15.3$ for $\triangle \nu=-1$ 
	(F-test probability of $2.1\times 10^{-4}$). 	
Then,
	when we try various abundance tables 
	(e.g., \texttt{angr}, \citealt{Grevesse1989AIPC..183....1G};
	 \texttt{aspl}, \citealt{Asplund2009ARA&A..47..481A})
	 with the \texttt{phabs} absorption model \citep{Balucinska1992ApJ...400..699B},
	the same \texttt{comptt} continuum 
	changes $N_{\rm H}$ up to by $\sim$50\%.
Under these uncertainties,
	we assume $N_{\rm H}=1.7\times 10^{22}$\,cm$^{-2}$,
	derived from the best-fit \texttt{comptt} model,
	as the default value in the discussion\footnote{If the Fe-K$\alpha$ line is due to fluorescence from an optically thin
gas which spherically surrounds the source, its equivalent width is
expected to be linearly correlated with the column density,
$N_\mathrm{H}$ as
$\textrm{EW}=100(N_\mathrm{H}/10^{23}\,\mathrm{cm}^{-2})$
\citep{Inoue1985SSRv...40..317I}. The line could therefore be produced
in a gas with a column density of 
$N_\mathrm{H}=2.9 \times 10^{22}\mathrm{cm}^{-2}$.}.
Subtracting the 21\,cm interstellar column density in the direction to 
	4U~1954+319, $N_\mathrm{H}=0.89\times 10^{22}\,\mathrm{cm}^{-2}$ 
	\citep{Kalberla2005},
	the intrinsic column density in the binary system is 
	$N_\mathrm{H}\sim 0.81\times 10^{22}\,\mathrm{cm}^{-2}$.


We also search the HXD-PIN data for a  possible presence of a 
cyclotron resonance scattering feature (CRSF), which provides 
a direct probe of the $B$-field at the accretion column
\citep{Makishima1999ApJ...525..978M,Schonherr2007A&A...472..353S}. 
Although a shallow structure at
$\sim$40 keV is visible, the depth of the absorption
(\texttt{cyclabs}) was constrained to $D<1.1$ (1$\sigma$). 
Thus, no CRSF was found in the {\it Suzaku} data.

\subsection{X-ray spectrum in the 2011 quiescent state}
In the first {\it Suzaku} observation performed in 2011 
	before the flare,
	the source was fainter by more than an order of magnitude,
	with the absorbed 2--10\,keV 
	X-ray flux $5.36\pm0.03 \times 10^{-12}\,\mathrm{erg}\,\mathrm{s}^{-1}\,\mathrm{cm}^{-2}$
	(\texttt{comptt} model).
Since
the HXD-PIN detection is marginal, we only took the XIS-FI and BI
spectra into account for this data set.
Table~\ref{tab:TimeAveragedSpectra2012} summarizes the spectral models
used to fit the data. 
Simple power-law and blackbody models failed 
with $\chi^2_\nu>1.3$, while the \texttt{diskbb},
\texttt{cutoffpl}, \texttt{pow*highecut}, and \texttt{compTT} models
were all acceptable.
Using the same \texttt{comptt} model, 
the unabsorbed 1--10 keV flux was 
$6.38\pm0.09 \times 10^{-12}\,\mathrm{erg}\,\mathrm{s}^{-1}\,\mathrm{cm}^{-2}$.
The 1$\sigma$ upper-limit
intensity of a narrow ($\sigma=10$\,eV) Fe K$\alpha$ line at
6.4\,keV was $I_{\rm Fe}<9.8 \times 10^{-7}\,\mathrm{ph}\,\mathrm{cm}^{-2}\,\mathrm{s}^{-1}$.

\subsection{Luminosity and mass-accretion rate} 

As seen above, between the quiescent phase in 2011 and the outburst
phase of 2012 the time-averaged X-ray flux of 4U~1954+319 changed by a
factor of about 100. Assuming a distance of 1.7\,kpc, the fluxes observed 
in 2011 and 2012 correspond to 1--70\,keV luminosities of
0.022$L_{35}$ and 2.1$L_{35}$ with 
$L_{35}=10^{35}\,\mathrm{erg}\,\mathrm{s}^{-1}(d/1.7\,\mathrm{kpc})^2$.
Correcting for the X-ray absorption increases these luminosities 
to 0.025$L_{35}$ and 2.2$L_{35}$, 
respectively. These values fall in the luminosity range
typically seen in SyXBs, $L_\mathrm{x}\sim
10^{33}$--$10^{35}\,\mathrm{erg}\,\mathrm{s}^{-1}$
\citep{Lu2012MNRAS}.

We can estimate the  accretion rate onto the NS as
\begin{equation}
\dot{M}_\mathrm{NS}=\frac{R_\mathrm{NS}}{\eta GM_\mathrm{NS}} L_\mathrm{X}
        =\begin{cases}
4.7\times 10^{13}\,\mathrm{g}\,\mathrm{s}^{-1} & \text{in 2011}\\
3.8\times 10^{15}\,\mathrm{g}\,\mathrm{s}^{-1} & \text{in 2012}
\end{cases}
\end{equation}
where $\eta=0.3$ is the efficiency of accretion, 
$R_\mathrm{NS}=10$\,km is the NS radius, 
and $M_\mathrm{NS}=1.4 M_{\odot}$ is the NS mass.

\begin{figure*}
\includegraphics[height=6cm]{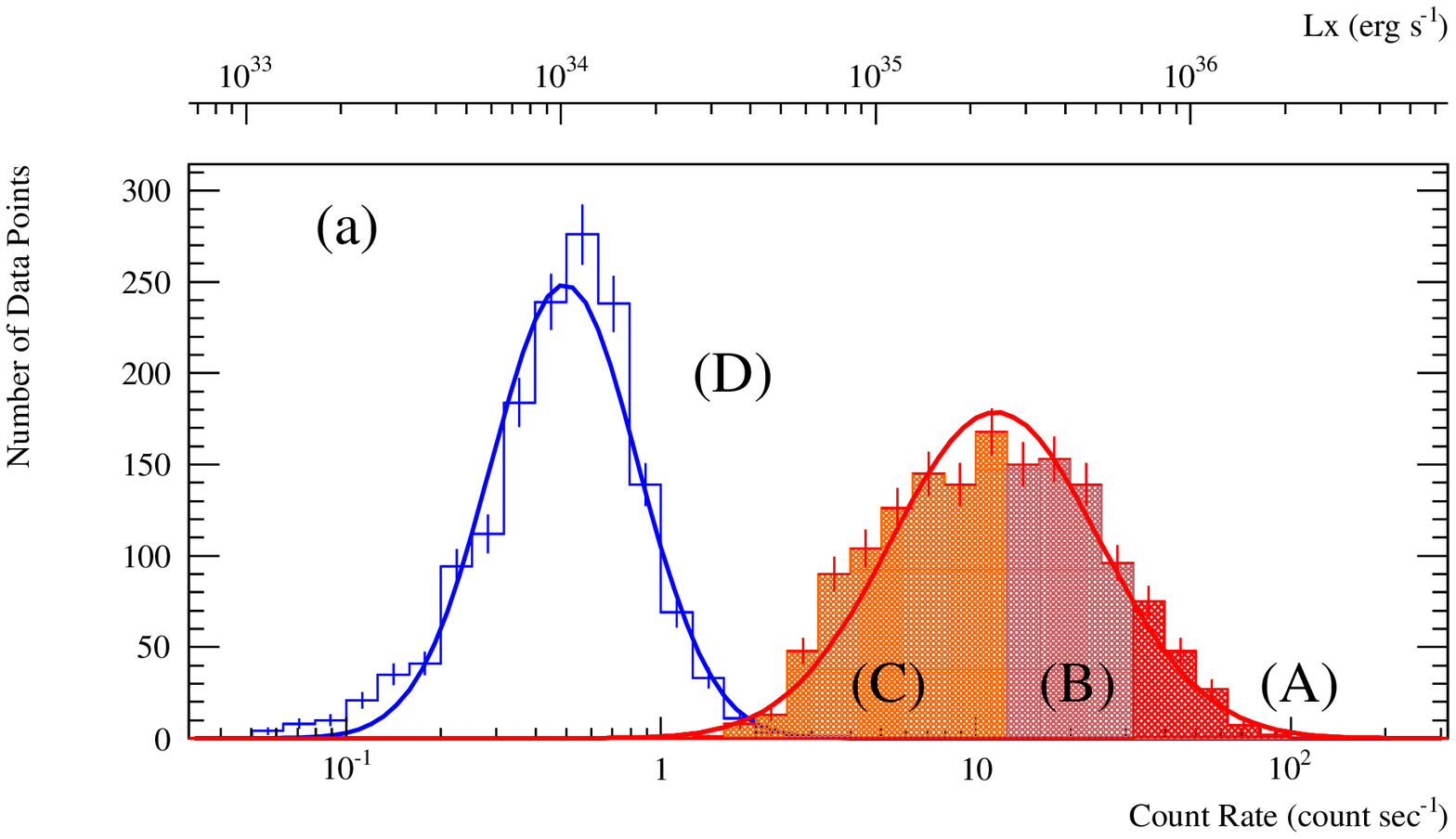}
\includegraphics[height=6cm]{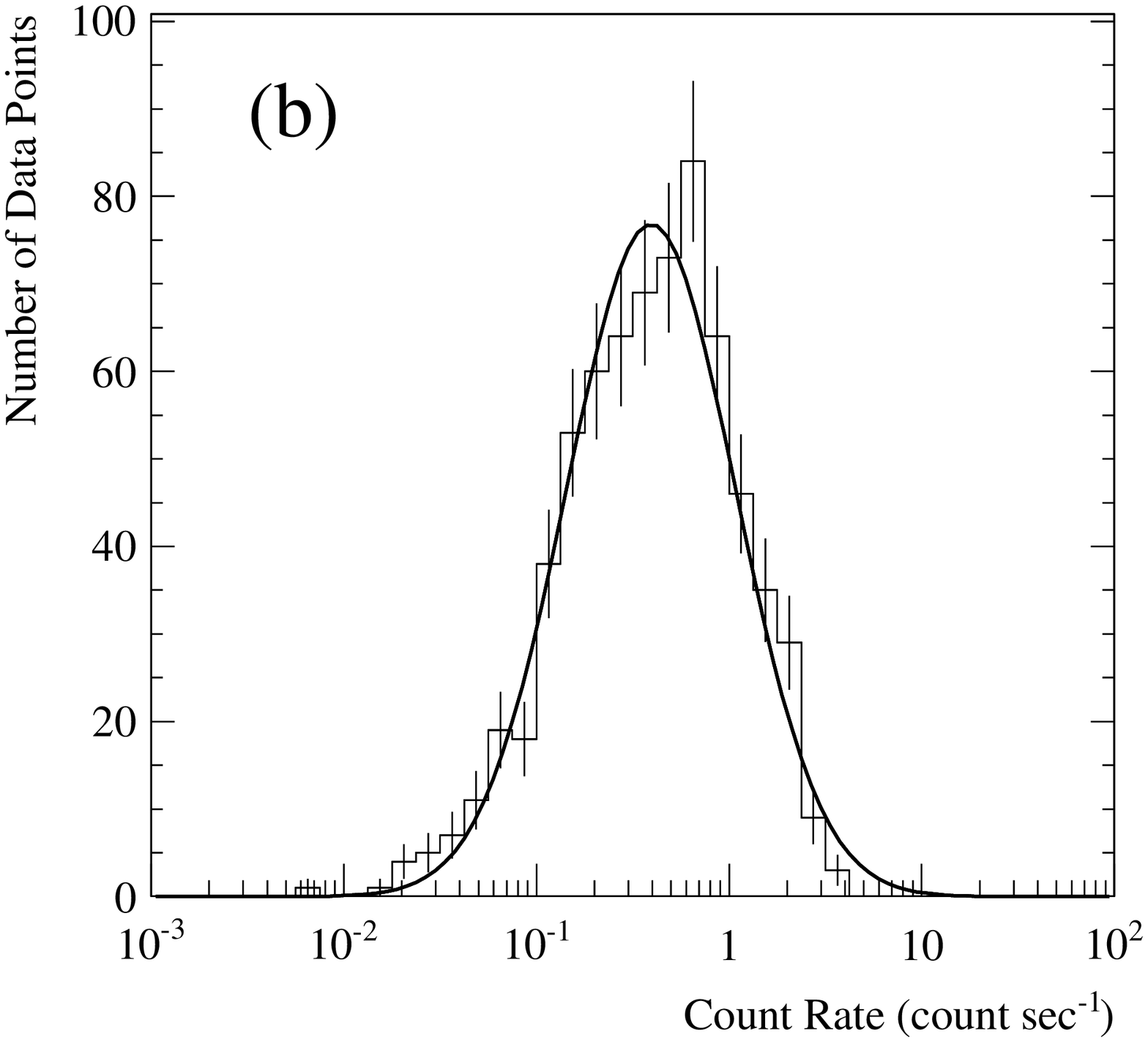}
\caption{ Histogram of count rates of the (a) 40\,s binned 0.5--10
  keV XIS and (b) 80-s binned 15--70 HXD-PIN light curve. Panel (a)
  shows the summed XIS rates in 2011 (blue) and 2012 (red). The
  corresponding luminosity is shown in the upper axis where the XIS
  0.5--10 keV rate is converted to the average 0.5--100\,keV X-ray
  luminosity using the 2012 observation and assuming $d=1.7$\,kpc. The
  solid curves in both panels show the best-fit Gaussians. Best fit
  Gaussian mean $p$ and standard deviation sigma $\sigma$ in log-space are derived to be
  $p=-0.31\pm0.01$, $\sigma=0.23\pm0.01$ (2011), $p=1.06\pm0.01$,
  $\sigma=3.31\pm0.01$ (2012) for panel (a), and $p=-0.41\pm0.02$,
  $\sigma=0.44\pm0.01$ for panel (b), respectively. The three regimes,
  bright (A; $>30\,\mathrm{cnt}\,\mathrm{s}^{-1}$), moderate (B;
  14--$30\,\mathrm{cnt}\,\mathrm{s}^{-1}$), dim phases (C;
  $<14\,\mathrm{cnt}\,\mathrm{s}^{-1}$) and the 2011 quiescent state
  (D; $\lesssim 2 \mathrm{cnt}\,\mathrm{s}^{-1}$) are shaded. }
\label{fig:log-normal_distribution}
\end{figure*}

\begin{figure*}
\includegraphics[height=10cm]{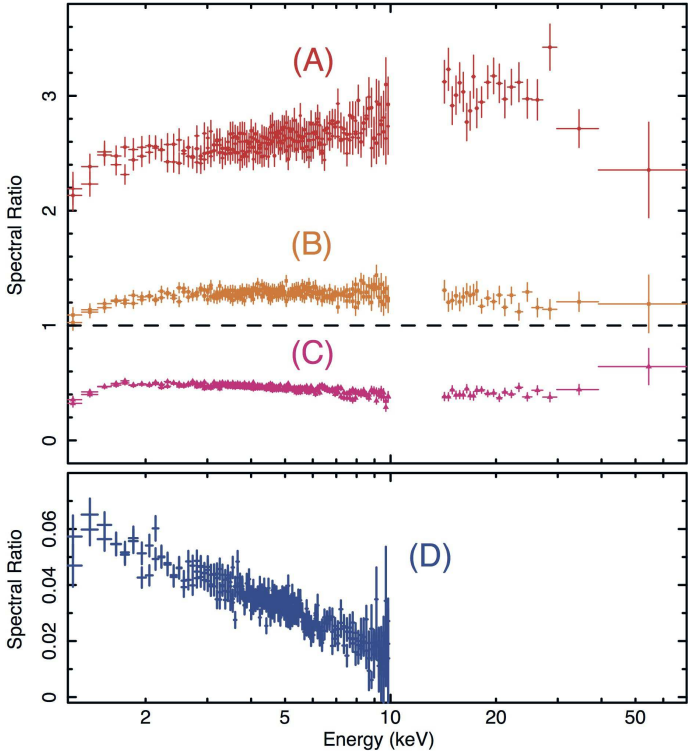}
\caption{ Spectral ratios of 4U~1954+319 to the average one in 2012.
  Four different X-ray luminosity states, (A), (B), (C), and (D) which
  were defined in Fig.~\ref{fig:log-normal_distribution}. }
\label{fig:spectral_ratio}
\end{figure*}

Since 4U~1954+319 is a highly variable source (\S\ref{Light curves and pulse  profiles}),
	we further investigated, in  Figure~\ref{fig:log-normal_distribution},
	distributions of  count rates in the XIS and HXD-PIN light curves,
	referring to Vela~X-1 studies by \citet{Furst2010A&A}.
The count rate distributions closely follow a log-normal distribution 
	in both the quiescent phase (2011, Fig.~\ref{fig:log-normal_distribution}a)
	 and the outburst (2012, Fig.~\ref{fig:log-normal_distribution}a and b).
Assuming the 2012 time-averaged spectral shape, 
	we further converted these to luminosity  
	at the top of Fig.~\ref{fig:log-normal_distribution}a.
As previously shown in Fig~\ref{fig:raw_spec},
	the spectrum becomes harder as the X-ray luminosity increases 
	from the quiescent phase (2011) to the outburst (2012).
Such a hardening trend is also detected within the 2012 observation
	when we divide the data into three intensity-sorted spectra
	(Fig.~\ref{fig:spectral_ratio}).

\section{Discussion}\label{discussion}
\subsection{Summary of the observations}
\label{Summary of the observations}
Combining the \textit{Swift}/BAT and \textit{RXTE}/ASM long-term
monitoring with the \textit{Suzaku} observations, 
we derived the following observational results 
which need to be explained:
\begin{enumerate}
\item The slow 5.4\,hour NS spin period of 4U~1954+319 exhibits a
  $\sim$7\% fluctuation over $\sim$8 years with at least four
  reversals in the sign of the period derivative.

\item In the 2011 October and 2012 November \textit{Suzaku} observations, 
  before and after the 2012 November X-ray outburst,
  the 1--70 keV X-ray luminosities were $2.2\times
  10^{33}\,\mathrm{erg}\,\mathrm{s}^{-1}$ and $2.1\times
  10^{35}\,\mathrm{erg}\,\mathrm{s}^{-1}$, respectively.
  In the former and the latter observations, 
  the source was in a spin-down and a spin-up phase, respectively.
  
\item The 0.5--70\,keV X-ray pulse profile shows a main peak and
  a sub peak. 
  The pulsed fraction mostly increases with energy from $\sim$60\% at 3\,keV
  to 80\% at 30\,keV.
  
\item During the 2012 outburst, the X-ray flux was highly variable on 10--1000 s
  time scales, with many irregular short flares each with a typical duration
   of $\lesssim$100\,s. The Fourier power
  density spectrum is dominated by red noise, 
  i.e., became redder from $-0.77$ in 2011 to $-1.66$ in 2012.
  The X-ray count rates follow a log-normal distribution in both
  observations.
  
\item The 0.8--70~keV phase-averaged spectra in the 2012 outburst can be
  represented by the Comptonization spectral models; \texttt{compTT},
  \texttt{BB+NPEX}, or \texttt{BB+highecut}. A fluorescent
  Fe-K$\alpha$ line at 6.37\,keV is detected with an equivalent width
  of $\mathrm{EW}=28.6\pm0.3\,\mathrm{eV}$, and a line flux of $I_{\rm
    Fe}=7.5^{+0.8}_{-0.7}\times
  10^{-5}\,\mathrm{photons}\,\mathrm{cm}^{-2}\,\mathrm{s}^{-1}$. The
  line is narrow, with a $1\sigma$ upper limit for the Gaussian width
  of 50\,eV. 
	The quiescent 2011 spectrum is also well described by 
	\texttt{comptt} with $kT\sim0.85$\, keV,
	or also by \texttt{diskbb}, \texttt{cutoffpl}, and \texttt{highecut} models.
  
\item The 2012 XIS spectrum is harder than that measured in 2011. The
  intensity-sorted X-ray spectra in 2012 slightly harden with source brightness. 

\item The 1--70\,keV band does not contain any statistically
  significant cyclotron lines.

\end{enumerate}

\subsection{Are SyXBs a subclass of LMXB?}

X-ray binaries hosting NSs are conventionally classified based on
stellar types of the optical counterparts. Commonly, HMXBs host a
strongly-magnetized NS and a high-mass companion (i.e., Be stars or OB
supergiants), 
while LMXBs
	\red{are thought to} host a weakly magnetized NS and 
	a low-mass star (e.g., K-type stars).
\red{
However,
	this old optical identification 
	ignores accretor properties (e.g., NS $B$-field) 
	which mainly determine the X-ray radiation.
In fact, 
	the conventional stereotype
	assuming weakly-magnetized NSs in LMXBs 
	is now challenged 
	by recent observational evidence of strongly magnetized NSs in LMXBs,
	e.g., 4U~1626-67 \citep{Camero-Arranz2012A&A} and 4U~1822-37 \citep{2013arXiv1311.4618S}.
	}

\begin{figure*}
\includegraphics[scale=0.6]{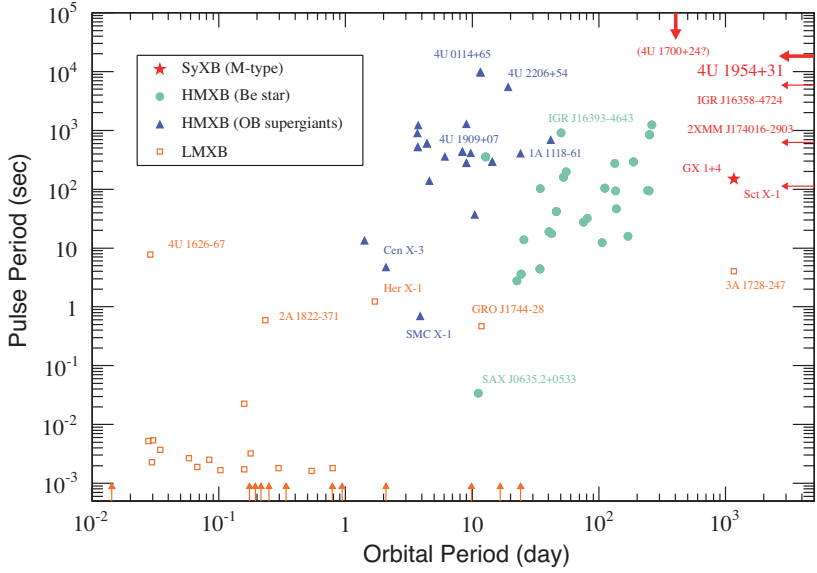}
\caption{ Known SyXBs (red stars) on the Corbet Diagram. Sources with
  either an orbital or rotation period detected are indicated as (red)
  arrows. HMXBs hosting a Be star (green filled circles), OB supergiant
  (blue filled triangles) and LMXBs (orange open squares; only X-ray
  pulsars) are also shown from the catalogs \citep{Liu2006A&A,
    Liu2007A&A, Bodaghee2007A&A...467..585B}. Arrows on the bottom
  axis represent atoll and Z-type LMXB systems for which only orbital
  periods were measured \citep{Liu2007A&A}. }
\label{fig:corbet}
\end{figure*}

The required overhaul of this outworn notion
 	is more clearly shown in SyXBs.
In contrast to the conventional classification
	which makes SyXBs the same category as LMXBs,
	the apparent high pulsed fraction (60--80\%) of SyXBs (Fig.~\ref{fig:pulse_profile},
	\S\ref{Summary of the observations}, point 3) 
	suggests a similarity to strongly-magnetized and highly-pulsed NSs 
	in HMXBs 
	(pulsed fraction $\sim$20--100\%, \citealt{Lutovinov2008AIPC.1054..191L})	
	rather than to non-pulsed NSs in LMXBs 
	or to weakly-pulsed millisecond X-ray pulsars
	(typical pulsed fraction at a few percent, \citealt{Patruno2012arXiv1206.2727P}).
In addition, if we look at the location of SyXBs in the ``Corbet
diagram'' \citep{Corbet1986MNRAS.220.1047C}
shown in Figure~\ref{fig:corbet}, SyXBs again resemble HMXBs due to
their wide-orbit and slow NS rotation. Finally, SyXB X-ray spectra,
and that of 4U~1954+319 in particular, and the other behavior
summarized in \S\ref{Summary of the observations} also suggest a
closer similarity of 4U~1954+319 to X-ray pulsars in HMXBs with a
$10^{12}$--$10^{13}$\,G $B$-field.


Even though many of these points argue towards a closer similarity of SyXB and HMXB, 
\red{such an association}
would be a significant challenge to our current
understanding of X-ray binary evolution. 
Stellar evolution theory shows that the evolutionary time scale for the companion of the
NS to become an M4-5 III counterpart is much longer than the
typical lifetime of binary systems with OB-type donors,
and thus, the 4U~1954+319 system would be old.
If one dogmatically expected NS $B$-field decay,
	the high $B$-field indicated by the strong X-ray pulsation
	easily causes a contradiction to the assumption of an old binary system.
Furthermore, 
	it is also difficult to keep such a binary system after a supernovae explosion,
	since
	the common envelop hypothesis, 
	which is usually assumed to make close binary systems hosting NSs in LMXBs,
	cannot be reconciled with the suggested large orbit of SyXBs (Fig.~\ref{fig:corbet}).
Te conventional evolutional theory leaves us with a real conundrum.

In addition to needing to study 
the population synthesis of such binaries
\citep[see,
e.g.,][]{Lu2012MNRAS,Chakrabarty1997ApJ}, 
the observed nature of SyXB requires us to investigate more exotic scenarios:
e.g., 
a magnetar-descendent was captured by an evolved M-type giant	
	in a close encounter 
	via a magnetic braking interaction 
	between the NS strong $B$-field and a large atmosphere of the M-type giant.
An alternative possibility is 
	an accretion-induced collapse (AIC) of a white dwarf 
	\citep{Nomoto1991ApJ...367L..19N}
	which potentially produces not only weak-field pulsars
	but also a magnetar \citep{Thompson1995ASPC...72..301T}.	

In the following, we take the features of 4U~1954+319 listed in
\S\ref{Summary of the observations} as observational clues to the
mystery of SyXB, comparing the conventional accretion model and the
revisited quasi-spherical accretion model, regardless of the
conventional HMXB and LMXB classification 
and population synthesis.

\subsection{The X-ray Spectrum of 4U~1954+319}\label{discuss:comp}

\begin{figure*}
\includegraphics[scale=0.36]{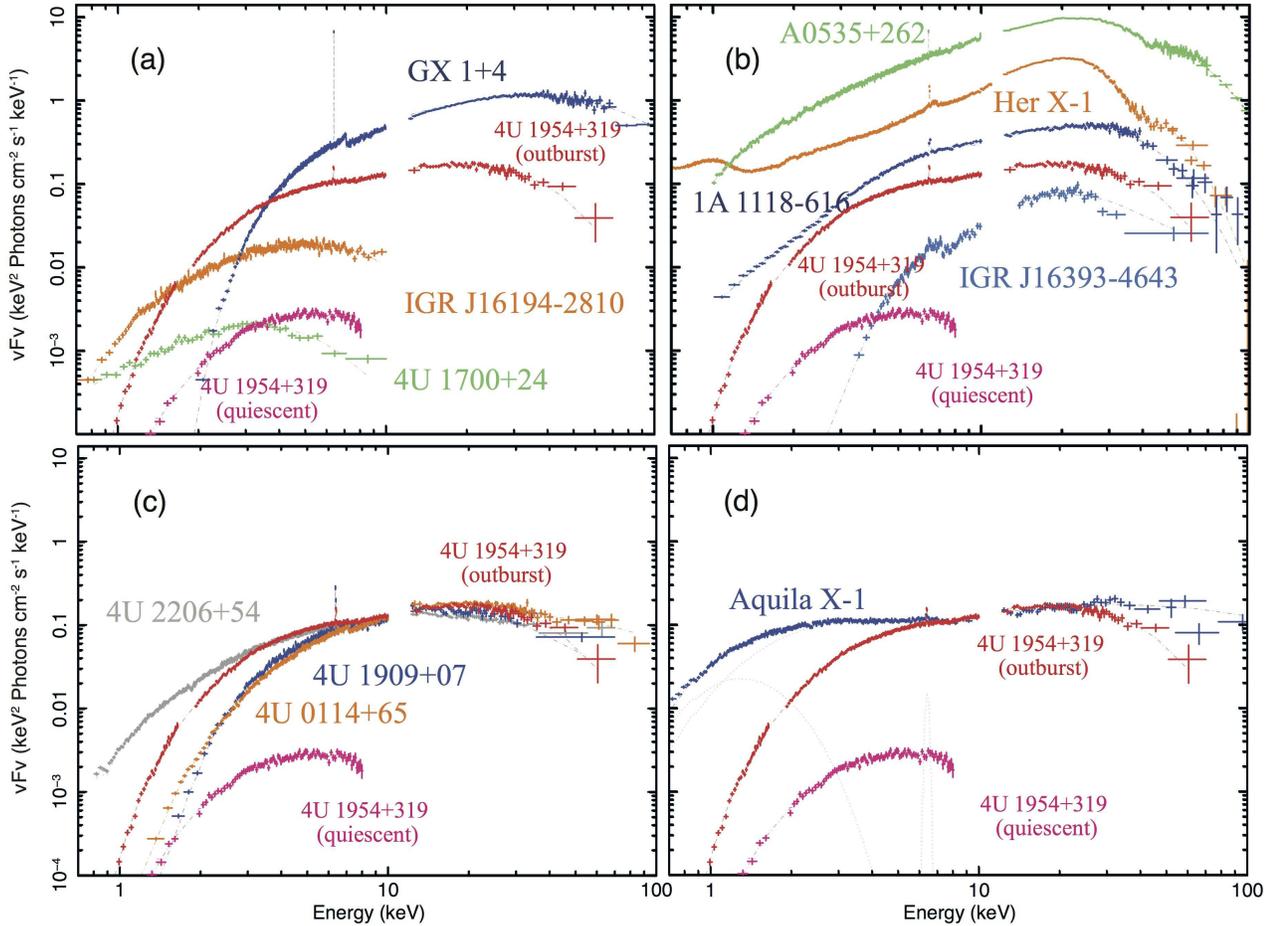}
\caption{ Spectral comparison of 4U~1954+319 with different categories
  of NS binaries: (a) other SyXBs, (b) CRSF sources (c) long period
  pulsars and (d) the LMXB Aquila~X-1  \citep{Sakurai2012PASJ} in the low/hard state. }
\label{fig:spec_comp_with_others}
\end{figure*}

The first clues on the nature of 4U~1954+319 come from
	broad-band spectroscopy. 
We assume that 	
	the observed X-ray spectrum is uniquely characterized 
	mainly by $\dot{M}$, $B$-field, and inclination of the compact object 
	regardless of the type of the mass-donor.
As shown in \S\ref{sec:analysis} and previously suggested by
\cite{Masetti2007A&A...464..277M}, spectral modeling implies that
Comptonization is the dominant process forming the X-ray continuum,
similar to the accretion column of the canonical X-ray pulsars in
HMXBs\footnote{Note that previous observations also included a soft
  excess which could be described with plasma emission with
  $kT\sim50$\,eV \citep{Masetti2007A&A...464..277M} 
  which is not seen here since the {\it Suzaku} low energy data below $\sim$1 keV 
  have uncertainties due to the contamination on the optical blocking filter.}. 	
Thus,	
	in Figure~\ref{fig:spec_comp_with_others}, 
	we now compare
	the X-ray spectrum of 4U~1954+319 
	with different X-ray spectral shapes of other related X-ray pulsars.

Figure~\ref{fig:spec_comp_with_others}a shows a comparison of the
X-ray spectrum of 4U~1954+319 with all archival \textit{Suzaku} SyXB
observations:
GX~1+4, 
4U~1700$+$24 \citep{Nagae2008}, 
and IGR~J16194$-$2810 \citep{Kitamura2013PASJ}. 
The rotational, orbital, and optical information of these X-ray sources
is listed in Table~\ref{tab:SpectralComparison} with the
\textit{Suzaku} observation records and references. To extract these
spectra we use the same procedure as for 4U~1954+319
(\S\ref{data_processing}). The $\nu F_{\nu}$ spectra were obtained by
fitting the data with a \texttt{cutoffpl} model with interstellar
absorption (\texttt{phabs}), and, if needed, an iron line. 
GX~1$+$4 exhibits harder spectra peaking at $\sim$20--40 keV
than 4U~1954+319. The spectral shape of IGR~J16194$-$2810, for which
orbital and pulse periods have not yet been detected, is similar to
that of 4U~1954+319 in quiescence, and thus their accretion environment
is expected to be similar.
        
In many accreting NSs with $B$-fields in the $10^{12}$\,G
regime, cyclotron lines at energies $E_{\rm
  cyc}=11.6(B_\mathrm{cyc}/10^{12} \textrm{G})(1+z)^{-1}$\,keV have been seen, where
$z\sim 0.2$ is the gravitational red shift of the NS 
\citep{caballero2012,Schonherr2007A&A...472..353S}. In
Fig.~\ref{fig:spec_comp_with_others}b we compare 4U~1954+319 with
well-studied CRSF sources. The 1.24\,s X-ray pulsar Her~X-1 is a
prototypical CRSF source with $E_\mathrm{cyc}$$\sim$$36$\,keV ($B\sim
3.1\times 10^{12}$\,G) with an unusual optical companion 
(A9--B companion with a mass of $M_{\rm c}\sim 2.3M_{\odot}$). 
The HMXBs 1A~1118$-$616 and A~0535+262
exhibit the strongest $B$-fields among known CRSF sources,
reaching $B\sim 4.7\times 10^{12}$\,G and $\sim$$4.1\times
10^{12}$\,G, respectively. 
IGR~J16393$-$4643 was originally classified as a SyXB,
	but finally identified as a HMXB \citep{Bodaghee2012ApJ...751..113B, Pearlman2011HEAD...12.4206P}.
As shown in
Fig.~\ref{fig:spec_comp_with_others}b, the spectral shape of
4U~1954+319 is similar to 1A~1118$-$616 and A~0535$+$262
rather than Her X-1. Although a widely accepted physical model has not
yet been established to explain the X-ray continuum, it has been
suggested that due to cyclotron cooling the high energy cutoff appears
around the CRSF energy. This means that the X-ray continuum is
expected to extend to higher energies as the $B$-field becomes
stronger \citep{Makishima1999ApJ...525..978M}.

The non-detection of a CRSF in 4U~1954+319 with the XIS and HXD means
that either the $B$-field is not in the range $B=9.0\times
10^{10}$--$6.0\times 10^{12}$ G, or that the cyclotron line is too
faint to be detected in the data. The latter is not too unlikely, with
a number of sources with similar behavior being observed. A possible
explanation for this is that CRSFs can be smeared out and/or filled by
photon spawning, i.e., the emission of photons close to the resonance
energy during their de-excitation from higher Landau levels
\citep{Schonherr2007A&A...472..353S, Furst2011_4U1909}. If, on the
other hand, $B\gtrsim 10^{13}$\,G, according to
\citet{Makishima1999ApJ...525..978M} we would expect a harder X-ray
continuum than that of HMXB with a cyclotron line.

In Fig.~\ref{fig:spec_comp_with_others}c we compare the
spectrum of 4U~1954+319 with those of some HMXBs which have not shown
a clear CRSF: HMXB 4U1909+07, 4U~2206+54, and 4U~0114+65. The
phase-averaged spectral shapes of these sources are quite similar to
that of 4U~1954+319. 
From a spectral view point they can be regarded as triplets. 
4U~1909+07 is a HMXB with an OB companion hosting a 604-sec
X-ray pulsar in a 4.4\,d orbit. Its pulse period is reported to show a
random walk like behavior \citep{Furst2011_4U1909}, which also
resembles the period history of 4U~1954+319, and possibly suggesting
wind type accretion rather than the stable accretion disk. The other
two X-ray pulsars, 4U~2206+54 and 4U~0114+65, exhibit very long
rotational periods of $\sim$5500\,s and $\sim$9700\,s, respectively 
	\citep{Finley1992A&A...262L..25F, Reig2009A&A...494.1073R,Masetti2006A&A...445..653M}.
Thus, we expect that such a long rotational period is strongly related to
	the spectral shape in Fig.~\ref{fig:spec_comp_with_others}c of these sources. 

Let us finally compare, in Fig.~\ref{fig:spec_comp_with_others}d,
	the spectrum with the low luminosity
	($L_\mathrm{X}\lesssim 10^{36}\,\mathrm{erg}\,\mathrm{s}^{-1}$)
	low/hard state of LMXBs as 
	motivated by \citet{Kitamura2013PASJ}
	who tried to search for similarities between 
	SyXBs and low-field LMXB-NSs (e.g., Aql~X-1)
	rather than high-field NSs,
	based on the optical classification.
Although the $B$-field of such LMXBs starts to dominate the accretion flow
	and an optically thin spherical flow would realize a similar geometry as  the SyXB,
	as we already discussed,
	 SyXBs should be different from other LMXBs due to 
1) slow pulsation with a large pulsed fraction (Fig.~\ref{fig:folding}) suggesting a high $B$-field,
2) the large orbital period (Fig.~\ref{fig:corbet}) 
	suggesting wind-type accretion, 
3) the lack of clear evidence on the accretion disk, and 
4) no observations of LMXB-like strong soft blackbody spectra,
and 
5) the strong narrow 6.4 keV iron fluorescence line 
	which is more typically found in HMXBs 
	but rare ($\sim$10\%) in standard LMXBs \citep{Torrejon2010ApJ...715..947T}.
	In addition, 
	LMXBs are usually characterized by highly ionized 6.6--6.9 keV lines. 

In conclusion, the broad-band  4U~1954+319 spectrum best resembles 
	that of X-ray pulsars in HMXBs,
	especially slowly rotating pulsars (e.g., 4U~0114+65 or 4U1909+07) 
	without an apparent accretion disk 
	as well as another quiescent SyXB, IGR J16194$-$2810. 
Therefore, 
	a strong field $\gtrsim 10^{12}$ or higher is favored 
	rather than a low field as in non-pulsating or weakly-pulsed NSs in LMXBs.

\subsection{The long spin period of 4U~1954+319}

Let us now investigate the timing behavior of the source. 
As already discussed previously
\citep[e.g.,][and references therein]{Marcu2011ApJ,Corbet2008ApJ}, the
5.4\,h period of 4U~1954+319 can only be explained as being the rotation
period of the NS. Both an orbital modulation or an M-type
star pulsation, can be ruled out. The long-term pulse history implies
at least four spin-torque transitions between a spin-up and a
spin-down phase.
The period evolution \red{indicates} that outside the
outbursts the NS is gradually slowing down, while
during the outbursts it enters a spin-up trend correlated with high accretion rates. 
This behavior suggests that it is
reasonable to assume that the source is near an equilibrium period of
accretion momentum transfer.
        
\begin{figure*}
\includegraphics[scale=0.36]{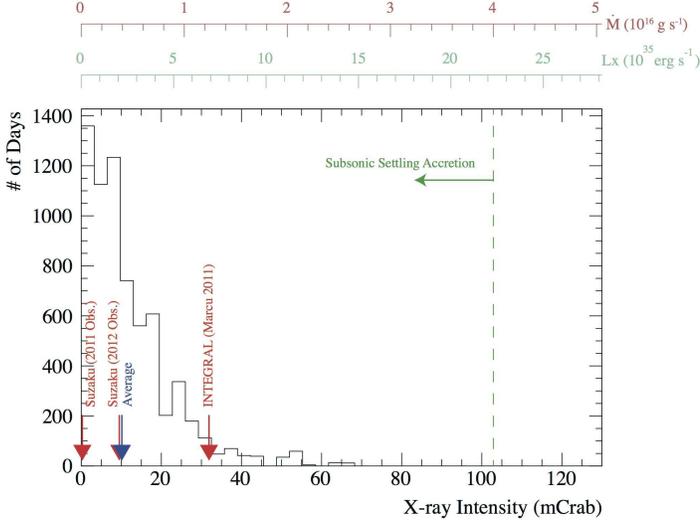}
\caption{ 
Histogram of the long-term 4U~1954+319 X-ray intensity 
	in mCrab unit derived from Fig.~\ref{fig:long-term_lc}.
Corresponding X-ray luminosity and mass accretion rate 
	are shown in the upper axes,
	assuming the distance at 1.7 kpc and $\eta$=0.3.
See Table~\ref{tab:Obslog} for the conversion factor from mCrab to X-ray flux.
The two {\it Suzaku} observations, 
	one {\it INTEGRAL} observation \citep{Marcu2011ApJ},
	and the average of the distributions are shown as arrows.
	}	
\label{fig:4u1954_luminosity_mdot_mCrab}
\end{figure*}

In the standard disk accretion model \citep{Ghosh1979ApJ}, the
equilibrium period is Eq. (76) in \cite{Shakura2012MNRAS}
\begin{equation}\label{eq:Ghosh}
  P_\mathrm{eq}\approx 7\mu_{30}^{6/7}\dot{M}_{16}^{-3/7} {\rm \ s},
\end{equation}
where $\dot{M}_{16}=\dot{M}/(10^{16}\,\mathrm{g}\,\mathrm{s}^{-1})$ is
the mass accretion rate, and 
$\mu_{30}=\mu/(10^{30}\,\mathrm{G}\,\mathrm{cm}^3)$ is the neutron
star's magnetic dipole moment. Inserting the values for 4U~1954+319
measured in 2012, i.e., 
$\dot{M}_{16}=0.38$ and $P_\mathrm{eq}\sim
19400$\,s ($\sim$5.4\,hour) yields an extremely high, magnetar-like
$B$-field reaching $B\sim 10^{16}$\,G.
This evaluation would hold even if considering the long-term history,
	since the average mass-accretion rate is close to 
	the value in the 2012 {\it Suzaku} observation.
This is clearly shown in Figure~\ref{fig:4u1954_luminosity_mdot_mCrab}
	where the long-term distribution of mass-accretion rates is illustrated 
	estimated from Fig.~\ref{fig:long-term_lc}.
	
While a magnetar-like $B$-field has been suggested for the SyXB
IGR~J16358$-$4726 \citep{Patel2007ApJ}, the field derived above is an
order of magnitude higher than the strongest magnetar field
\citep[SGR~1806$-$20, with
$2.1\times10^{15}\,\mathrm{G}$;][]{kouveliotou98a,Enoto2010ApJL}. 4U~1954+319 also
does not show any other magnetar-like properties, such as short
bursts or giant flares \citep{Enoto2009ApJ...693L.122E}. In addition, it is unclear whether standard
disk theory can be applied to SyXBs. For example,
\citet{Gonzalez2012A&A...537A..66G} show that the spin and X-ray
intensity behavior of the SyXB GX~1+4 cannot be explained either by
standard disk accretion theory or even by more advanced ideas such as
accretion from a retrograde disk. We therefore have to also consider 
alternative accretion models.

As the Corbet diagram (Fig.~\ref{fig:corbet}) implies large orbital
periods of SyXBs (e.g., $P_\mathrm{orb}=1161$\,d for GX~1+4;
\citealt{Hinkle2006ApJ}), it is reasonable to assume that 4U~1954+319
does not accrete via Roche lobe overflow but is a wind-fed accretor.
The large orbital period leads to a slow NS orbital velocity\footnote{
Although the orbital radius of 4U 1954+319 has not yet been detected,
	assuming a circular orbit with its orbital period at 500 d,
	we will derive the semimajor axis at $3\times 10^{13}$ cm
	via the Kepler's law and masses of the NS and M giant. 
Then the orbital velocity becomes $v=\sqrt{GM/a}\sim$30 km s$^{-1}$.
}.
Combining this  
	with a slow stellar wind compared to HMXBs 
	and 
	with low mass-accretion rate, 
	small angular momentum differences 
	are an intrinsic importance of the SyXBs 
	which makes the mass-accretion more quasi-spherical.

\begin{figure*}
\includegraphics[scale=0.3]{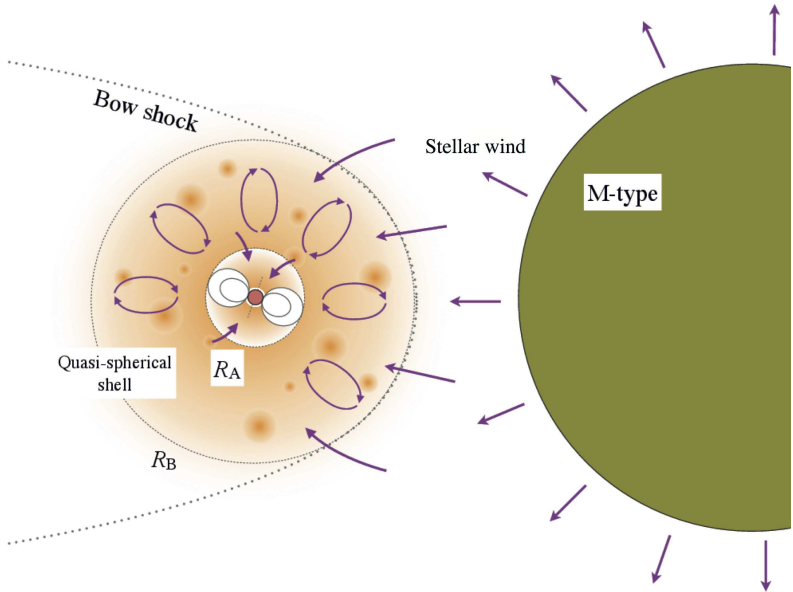}
\caption{ Schematic view of the quasi-spherical accretion and subsonic
  settling accretion (see details in \citealt{2013arXiv1307.3029S}). 
  $R_A$ and $R_B$ are the Alfv\'en and Bondi radii, respectively.
  }
\label{fig:schematic_view}
\end{figure*}

As recently shown by \citet{2013arXiv1302.0500S}, 
the X-ray luminosity of 4U~1954+319 actually puts the source in the regime of quasi-spherical,
subsonic setting accretion in the magnetosphere.
Figure~\ref{fig:schematic_view} shows a sketch of this alternative
wind accretion scenario for 4U~1954+319, which is possible for
$L_\mathrm{X} \le L_{\rm th}=4\times
10^{36}\,\mathrm{erg}\,\mathrm{s}^{-1}$ and $\dot{M}_{16} \le 4$.
Similar to the standard Bondi-Hoyle-Lyttleton accretion, in this scenario
a bow shock is formed in the wind at a characteristic distance ($R_\mathrm{B}$)
from the NS and matter entering this region
is accreted onto the compact object. 
Unlike the classical Bondi-Hoyle-Lyttleton accretion 
	at higher luminosities, 
	for the low $L_\mathrm{X}$ of 4U~1954+319, Compton cooling becomes
ineffective and the matter can settle in the magnetosphere with a
subsonic velocity through a quasi-spherical hot shell
(\S\ref{Introduction}). 
If this happens, the angular momentum can be
transferred via large-scale convective motions or turbulence in this
quasi-static shell. This results in an equilibrium spin period of
\citep{2012arXiv1212.2841P}
\begin{equation} \label{eq:equilibrium_period_spherical}
  P_\mathrm{eq}\simeq 13000\ \mu_{30}^{12/11} \dot{M}_{16}^{-4/11} v_8^4
  \left(\frac{P_\mathrm{orb}}{100\,\mathrm{d}}\right)\,\mathrm{s},
\end{equation}  
where $v_8=v/(1000\,\mathrm{km}\,\mathrm{s}^{-1}$) is the relative
velocity between the stellar wind and the NS.

The most uncertain parameter in
Eq.~(\ref{eq:equilibrium_period_spherical}) is $v_8$. The wind
velocity strongly affects $P_\mathrm{eq}$, 
but it is not well known for the late-type M giants due to lack of observations. Existing
measurements for wind speeds are mainly available for massive OB or
Wolf-Rayet stars with high mass-loss rates
($\dot{M}_\mathrm{wind}=10^{-10}$--$10^{-5}\,M_{\odot}\,\mathrm{yr}^{-1}$).
These radiatively-driven winds are 
	very fast (up to $\sim 2500\,\mathrm{km}\,\mathrm{s}^{-1}$; 
	\citealt{Lamers1999isw..book.....L}). 
The situation is different for SyXB
systems. Here no strong evidence for fast winds exists. Late F--M
stars are usually thought to have smaller mass-loss rates 
	($\lesssim 10^{-10}\,M_{\odot}\,\mathrm{yr}^{-1}$) with much slower terminal
velocities \citep[$\lesssim
100\,\mathrm{km}\,\mathrm{s}^{-1}$;][]{2008ASPC..401..166E}. These
slower speeds are consistent with the lower escape velocities of the
stars. For example, if the donor in 4U~1954+31 has $1.2\,M_\odot$ and a
radius of $80R_\odot$ as suggested by \cite{Mattana2006A&A}, then the
escape velocity from the companion star would be only
$\sim$$76\,\mathrm{km}\,\mathrm{s}^{-1}$. To our knowledge, the only
direct measurement of the wind speed in a SyXB was performed for the
M6-giant V2116 Oph, the donor star of the SyXB GX~1+4, where the wind
speed was found to be $\sim$100--200$\,\mathrm{km}\,\mathrm{s}^{-1}$
\citep{Chakrabarty1998ApJ...497L..39C,Hinkle2006ApJ}.

Assuming that the wind in 4U~1954+319 is comparable to that in GX 1+4,
Figure~\ref{fig:theory1} shows the dependence of the equilibrium period
for several plausible different combinations of $v_8$,
$P_\mathrm{orb}$, $\dot{M}$, and $B$ as calculated with
Eq.~\eqref{eq:equilibrium_period_spherical}. 
If 4U~1954+319 has a typical accreting NS 
	with a $B$-field of $10^{12-13}$\,G, 
	then its equilibrium period can only be explained by
	a fast M-star wind (e.g., $\sim$300 km s$^{-1}$, 
	probably even faster than that observed in GX~1+4)
	and the long orbital period (e.g., $\sim$500 d).

\subsection{Time scales of Irregular flares}

\begin{figure*}
\includegraphics[height=10cm]{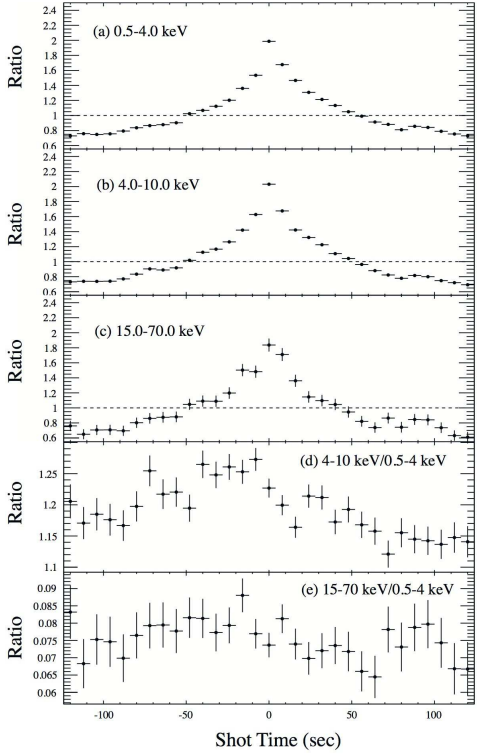}
\caption{ Background-subtracted stacked shot profiles of 4U~1954+31 in
  (a) the 0.5--4.0\,keV, (b) 4.0--10.0\,keV, and (c) 15.0--70.0\,keV
  bands. Shot peaks were determined from local maxima within
  $\pm$200\,s duration and with a threshold factor of $f=1.8$
  \citep{2013ApJ...767L..34Y}. The profiles in panels a, b, and c are
  normalized by the average rate in individual energy bands. (d)(e)
  Panels (d) and (e) show 
  the divisions of the profiles in panels (b) and (c) by that in 
  panel (a).
}\label{fig:shot_profile}
\end{figure*}

A further handle on the accretion mechanism at work in 4U~1954+319
comes from the short term variability of the source. The quiescent
state in 2011 shows a steady weak accretion (in
Fig.~\ref{fig:suzaku_lc_enlarge} a,b), while the outburst in 2012
exhibits many irregular flare on time scales of 
$\lesssim$100 s (Figs.~\ref{fig:suzaku_lc} and
\ref{fig:suzaku_lc_enlarge} c, d). This highly variable fluctuation
indicates a blobby mass accretion (Fig.~\ref{fig:schematic_view}).
The X-ray light curves show that the X-ray variation is
mainly due to short-time scale spiky flares. Similar spike-like
structures, also called ``shots'', were studied previously in analyses
of the light curves of accreting black holes such as Cygnus~X-1
\citep{Negoro1994ApJ...423L.127N,2013ApJ...767L..34Y}. In these
objects the spectrum gradually softens during the peak of a shot,
followed by an immediate hardening. Figure~\ref{fig:shot_profile}
shows a similar analysis for 4U~1954+319. 
The shots in the light curves are represented by a shape with nearly symmetric rise and
decay \red{(Fig.~\ref{fig:shot_profile} a, b, c)} 
on a timescale of $\sim$50\,s (at which the count rate drops 
below the average) with signs of a slight hardening 
in the early phase of the shot-like behavior 
	which is indicated in  \red{(Fig.~\ref{fig:shot_profile} d, e)}. 
	
\begin{figure*}
\includegraphics[width=0.5\textwidth]{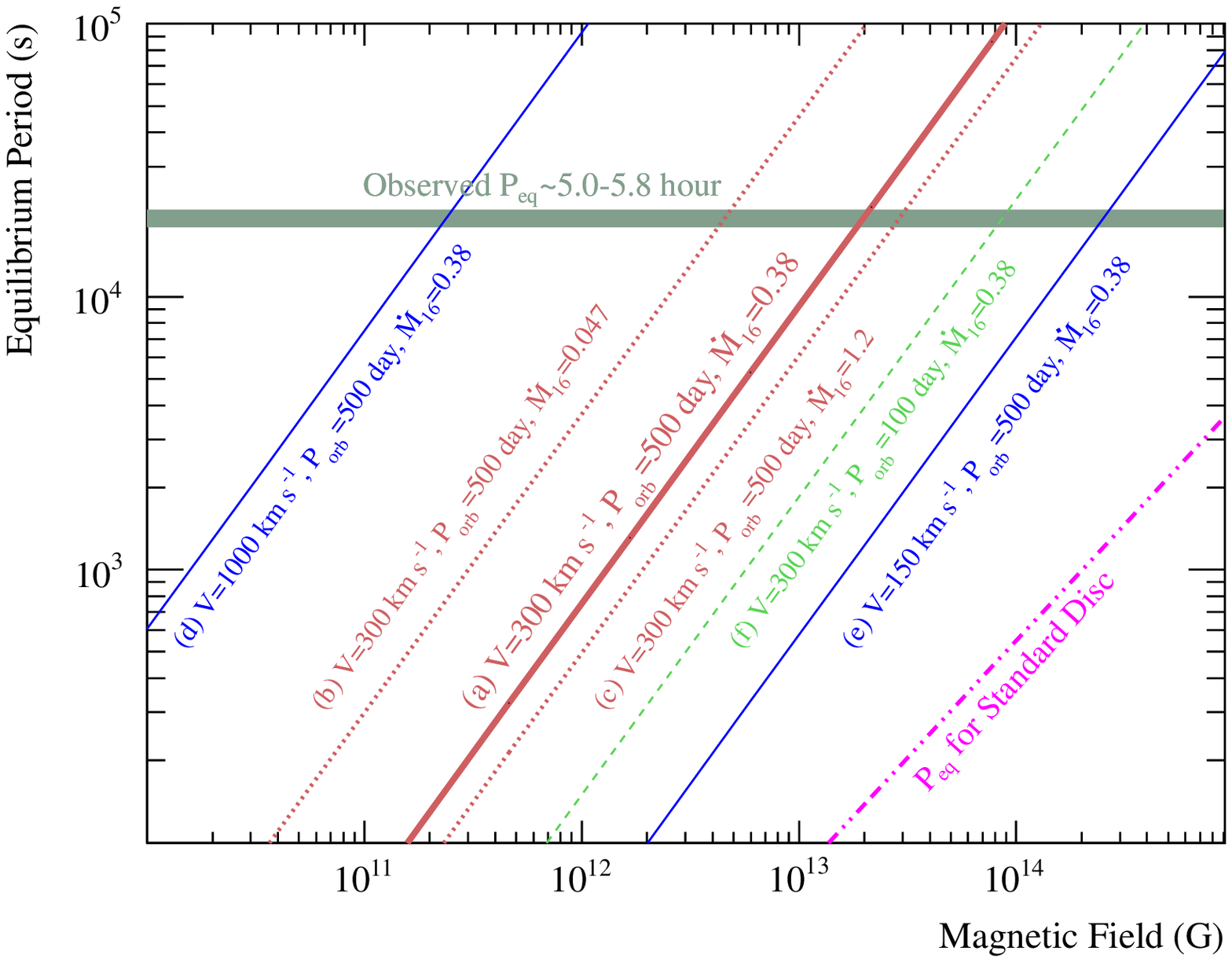}
\includegraphics[width=0.5\textwidth]{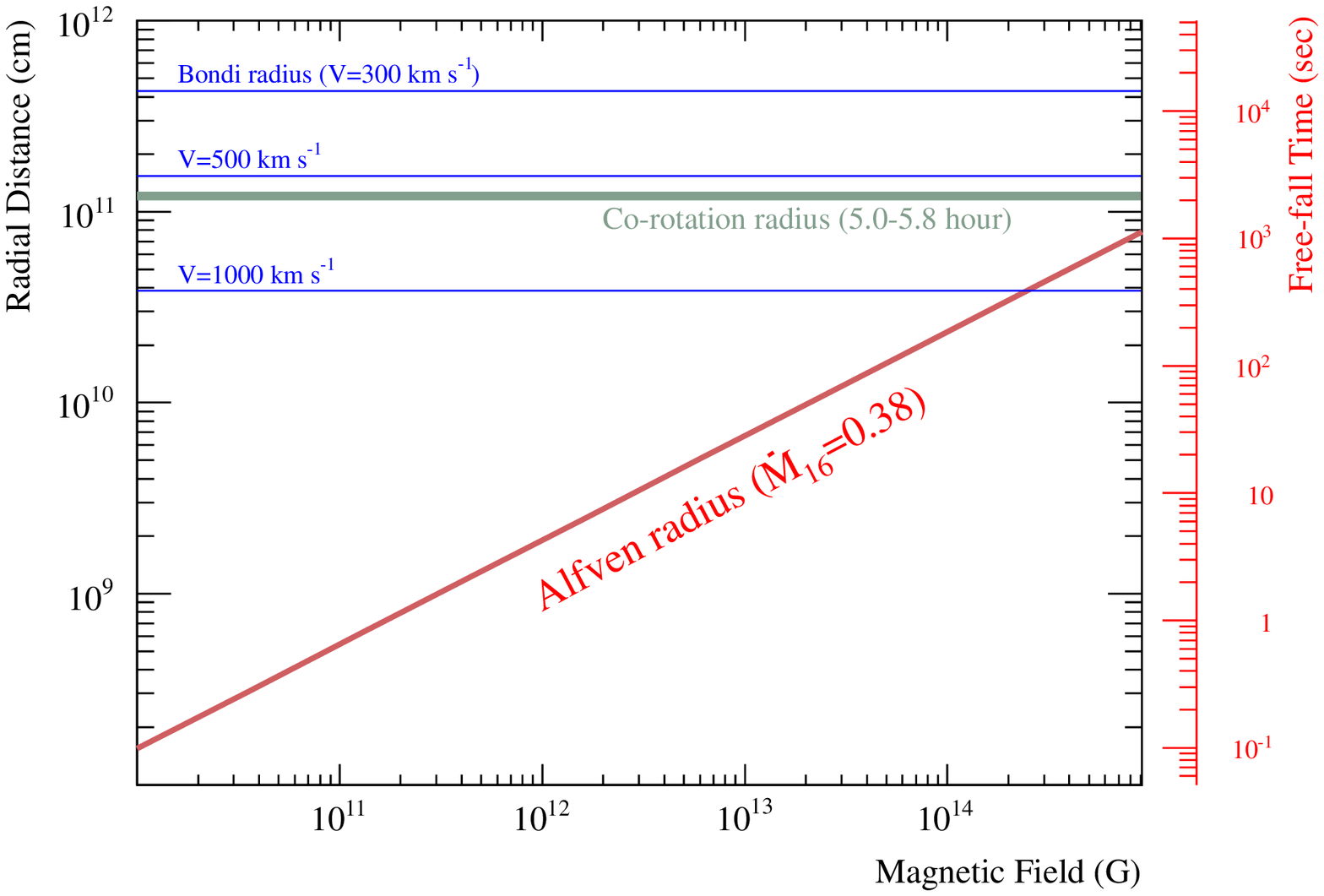}
\caption{Left: 
  The NS equilibrium period $P_\mathrm{eq}$
  as a function of the magnetic field $B$ and different stellar wind
  parameters (velocity, mass-accretion rate, and fixed $P_{\rm orb}$ at 500 d)
according to the quasi-spherical accretion model 
 \citep[][see Eq.~\ref{eq:equilibrium_period_spherical}]{2012arXiv1212.2841P}.
  The green box in the background shows the observed equilibrium
  period of 4U~1954+319, 5.0--5.8\,hours.  
  The magenta line at high $B$-fields shows the predicted equilibrium
  period for a standard disk \citep[Eq.~\ref{eq:Ghosh}
  and][]{Ghosh1979ApJ}. Right: The Alfv\'en radius from Eq.~\ref{eq:RA} as
  a function of the $B$-field strength and Bondi radii (Eq.~\ref{eq:RB})
  for different wind speeds (1000, 500, and 300 km s$^{-1}$), 
  assuming the observed $\dot{M}_{16}=1.6$.
  The green box shows the co-rotation radius. 
  The right hand axis shows the free-fall time onto the neutron star
  from each radius.}\label{fig:theory1}
\end{figure*}

A possible explanation for such shots 
	in 4U~1954+319 are individual blobby structures in the accretion flow. 
To see whether this interpretation is reasonable, we need to take a look at
the typical size and time scales of this system
(see Fig.~\ref{fig:theory1} right panel for a visualization).
In the model by \citet{2012arXiv1212.2841P} the stellar wind is
gravitationally captured by the NS inside the Bondi radius,
\begin{equation}\label{eq:RB}
  R_\mathrm{B}=\frac{2GM_\mathrm{NS}}{v^2}=3.78\times
  10^{10}\,v_8^{-2}\, \mathrm{cm}.
\end{equation}
For $v_8=0.3$, i.e., a stellar wind velocity of
$300\,\mathrm{km}\,\mathrm{s}^{-1}$, 
the Bondi radius $R_\mathrm{B}\sim 4.2\times 10^{11}$\,cm. 
Inside this radius a quasi-static shell is formed
above the magnetosphere.
This is the reservoir of matter from which
the NS accretes. 
In this model the co-rotation radius is
\citep{2012arXiv1212.2841P}
\begin{eqnarray}\label{eq:Rc}
  R_\mathrm{c}=\left(\frac{GMP^2}{4\pi^2}\right)^{1/3} 
  =1.21\times 10^{11}\,\mathrm{cm} \left( \frac{P}{5.40\,\mathrm{h}}\right)^{2/3}.
\end{eqnarray}
Even though the co-rotation radius $R_\mathrm{c}$ is very large, 
	it is still inside the Bondi radius if the stellar wind is slower than
$\sim$$600\,\mathrm{km}\,\mathrm{s}^{-1}$ (Fig.~\ref{fig:theory1} right).
This sets an upper limit for the wind speed in the system.

The size of the magnetosphere, characterized by the Alfv\'en radius where
the gas pressure becomes equal to the magnetic pressure
\citep{Ghosh1979ApJ}, is modified for quasi-spherical accretion at low
mass accretion rates and becomes \citep{2012arXiv1212.2841P,
  2013arXiv1302.0500S}
\begin{equation}
R_\mathrm{A}\sim 1.6\times
10^9\,\mathrm{cm}\left(\frac{\mu_{30}^3}{\dot{M}_{16}} \right)^{2/11}. 
\label{eq:RA}
\end{equation}
With $\dot{M}_{16}=0.38$ we find $R_\mathrm{A}=1.9\times 10^9$\,cm for
$B=10^{12}$\,G, and $6.7\times 10^9$\,cm for $B=10^{13}$ G,
respectively. This means that the Alfv\'en radius is smaller by 1--2
orders of magnitudes than $R_\mathrm{c}$ and $R_\mathrm{B}$.

This estimate for the size of the Alfv\'en radius is
roughly consistent with that found from spectral analysis. While the
2012 spectrum is Comptonized, during the 2011 quiescent state
Comptonization does not play a strong role. Approximating the
quiescent spectrum by a blackbody and using the Stefan-Boltzmann law,
the hot spot size on the NS is
$R_\mathrm{hot}=77\,\mathrm{m}\cdot(d/1.7\,\mathrm{\ kpc})$. Since the
hot spot size is determined by the location where matter couples onto
the $B$-field, we can use this measurement to estimate the size of the
Alfv\'en radius \citep{Lamb1973ApJ}, 
\begin{equation}
  R_\mathrm{A} \lesssim 10^5\,\mathrm{km} \left( \frac{R_\mathrm{hot}}{100\,\mathrm{m}} \right)^{-2} \sim 5.9\times 10^9\,\mathrm{cm}.
\end{equation}
The measured hot spot radius thus yields a $B$-field in the
$10^{12}$--$10^{13}$\,G range (Fig.~\ref{fig:theory1}).

Depending on the ability of the plasma to enter the magnetosphere,
matter from the magnetosphere funnels onto the NS poles along the
$B$-field lines. The free fall time is given by
\begin{equation}
t_\mathrm{ff} = 51\,\mathrm{s} \left(\frac{r}{10^{10}\,\mathrm{cm}}\right)^{3/2}.
\end{equation} 
The free-fall timescale from the Alfv\'en radius for $B\sim10^{13}$\,G
is $\sim$30\,s, close to the typical time scale of individual shots
(Fig.~\ref{fig:shot_profile}). Thus, it is possible to interpret the
shot-like flares as produced during the free-fall of the blobs from 
the shell above the magnetosphere. 
Even though the cooling time scale of the matter is
complicated when considering different cooling processes, if we assume
Compton cooling as an effective process \citep{Shakura2012MNRAS},
the cooling time, 
\begin{eqnarray}        
 t_{\rm cool}=
 10^3 {\rm \ s} \cdot  \frac{1}{\dot{M}_{16}} \left(\frac{r}{10^{10} {\rm \ cm}} \right)^2 
 \end{eqnarray}
 becomes $\sim$1200 s at $B=10^{13}$ G and $\dot{M}_{16}=0.38$. This
 is longer than the typical free-fall time ($t_{\rm cool} > t_{\rm
   ff}$) which satisfies the assumption of a low-luminosity
 quasi-spherical accretion.

The log-normal distribution of the X-ray light curve
count rates (Fig.~\ref{fig:log-normal_distribution}) is also
consistent with the picture of quasi-spherical magnetospheric
accretion. As suggested by \cite{Furst2010A&A}, such a log-normal
distribution implies that the underlying physical process should be
multiplicative rather than additive, since addition of many
independent emissions will produce a Gaussian distribution due to
the central limit theorem. Such a process could be induced via a
positive feedback loop, where blob accretion induces subsequent
accretion, e.g., via enhanced Compton cooling, potentially creating 
positive feedback until the matter at the boundary of the
magnetosphere is exhausted. Alternatively, the structures leading to a
log-normal distribution of the accretion rate could also be due to a
highly inhomogeneous and clumpy stellar wind from the M-giant
companion or due to turbulent processes in the spherical shell itself. 

\subsection{Absorption in the spherical shell}\label{sec:absorption}

Finally let us compare the above accretion scheme 
	with the absorption properties of the circumstellar material. 
These come from both the absorption modeling of the X-ray continuum 
	and the analysis of the narrow Fe K$\alpha$ line at 6.4 keV. 
The latter was detected with \textit{Suzaku} in the 2012 observation, and
has been previously reported with \textit{BeppoSAX}
($\mathrm{EW}=51\pm20$\,eV) and \textit{RXTE} ($48^{+13}_{-10}$\,eV;
\citealt{Masetti2007A&A...464..277M}). 
A narrow iron line feature is characteristic to 
	HMXB-NSs rather than LMXB-NSs.

As a plausibility check, 
	we assume that $N_\mathrm{H}$ is directly
related to the properties of the shell in the low-luminosity wind accretion model. 
Using the {\it Suzaku}-derived $N_\mathrm{H}$, the number
density ($n_{\rm H}$) inside the spherical shell ($R<R_\mathrm{B}$) becomes
$\average{n_\mathrm{H}} = N_\mathrm{H}/R_{\rm B}=1.9\times
10^{10}\,\mathrm{cm}^{-3}$ ($v_8=0.3$), corresponding to a mass
density of $\average{\rho}=\average{n_\mathrm{H}} m_\mathrm{p} \sim
3.2\times 10^{-14}\,\mathrm{g}\,\mathrm{cm}^{-3}$ where $m_\mathrm{p}$
is the proton mass. If the NS is fed by blobs of matter
which have the density of the shell at the Alf\'en radius, then in
order to reach the average mass accretion rate
$\average{\dot{M}_{16}}=4/3\pi R_\mathrm{blob}^3\rho/t_\mathrm{shot}$
where $R_\mathrm{blob}$ and $t_\mathrm{shot}$ 
are the typical blob size and shot time scale, 
we find $R_\mathrm{blob}\sim 8.0\times 10^9$\,cm
	with $\average{\dot{M}_{16}}=0.38$ and $t_\mathrm{shot}\sim50$\,s.
It is intriguing that this initial blob radius is
comparable to the Alfv\'en radius (e.g., $R_\mathrm{A}=6.8\times 10^9$
cm at $B=10^{13}$ G) and much smaller than the quasi-spherical shell
size $R_\mathrm{B}$. Thus, the shell can store enough mass to
generate the observed multiple flares and sustain accretion for longer time intervals.

Finally, as a cross check we note that the ionization parameter of the
shell at the Bondi radius is $\xi=L_\mathrm{X}/\average{n_\mathrm{H}}
r^2=2.1\times 10^4 (r/10^{10}\,\mathrm{cm})^{-2} \sim 11$, i.e., the
outer part of the shell is at most moderately ionized, making it possible 
to produce the 6.4\,keV line in the shell.

\section{Conclusion}
\label{conclusion}

We analyzed two {\it Suzaku} data sets, 
	obtained in 2011 and 2012,
	of the SyXB 4U~1954+319 
	consisting of the slowly rotating 5.4 h X-ray pulsar 
	and a late type primary star (M4-5 III). 
Our main results are as follows;

\begin{enumerate}
\item 
We reconfirmed the slow rotation period
	and the large period fluctuations.
We also identified several additional features of 4U~1954+319,
	including a hard continuum without a clear CRSF,
	a narrow Fe-K$_{\alpha}$ line,
	and sporadic flares.
The continuum was harder when the source was brighter.	

\item
From spectral and timing properties,
	4U~1954+319 is suggested to have a strong $B$-field of $\gtrsim 10^{12-13}$\,G.
This is supported 
	by the high pulsed fraction of 60--80\%,
	the slow rotation,
	the hard featureless continuum,
	and the narrow Fe-K$\alpha$ line.
	
\item 
The NS is expected to be close to equilibrium 
	because of the four spin transitions and large period fluctuations. 
Such an evolution can be explained by 
	a quasi-spherical accretion in a subsonic accretion regime
	which is considered to apply to 4U~1954+319 
	because of its low luminosity ($0.023-2.1\times 10^{35}$\,erg\,s$^{-1}$),
	presumably wide orbital period,
	and rather low wind velocity from the M-type giant. 
	
\item
Recurrent irregular flares during the outburst have a typical time scale of $\sim$50\,s,
	which is interpreted as intermittent accretion from the Alfv\'en  radius 
	of the $\sim 10^{13}$\,G NS.
The log-normal distribution of the light curve
	indicates an underlying multiplicative process 
	(e.g., positive feedback) in the accretion. 

\item
The extreme magnetar-like extreme $B$-field ($10^{16}$\,G) 	
	derived from the standard disk accretion in Eq.\,(\ref{eq:Ghosh})
	is not required 
	if we assume a low luminosity and quasi-spherical accretion.
	
\item
The presence of a system like 4U~1954+319
	which is very likely to involve a high-field NS,
	provides another example which urges us to challenge 
	the conventional classification
	of NS-binaries into LMXBs and HMXBs.
\end{enumerate}
                
\acknowledgements The authors would like to express their thanks to
\textit{Suzaku} team for their prompt observation during the 2012
flaring activity. TE was supported by JSPS KAKENHI, Grant-in-Aid for
JSPS Fellows, 24-3320. We thank Pranab Ghosh and Hiromitsu Takahashi
for useful discussions on this source, 
Kunugawa Tomoya and Kenta Hotokezaka
for comments on the binary evolution. 
We thank the Deutsches Zentrum f\"ur Luft- und Raumfahrt for partial
funding under DLR grant number 50 OR 1207.



\begin{deluxetable}{lcc}
\label{tab1}
\tablecaption{Log of {\it Suzaku} observations of 4U~1954+319. \label{tab:Obslog}}
\tablehead{
\colhead{} & \colhead{2011} & \colhead{2012} \\
\colhead{} & \colhead{AO6} &  \colhead{AO7 ToO} 
}
\startdata
Observation start & Oct. 23 08:05 & Nov. 1 02:39 \\
Start MJD & 55857.337 & 56232.110 \\
Observation end  &  Oct. 24 17:17 & Nov. 2 12:02 \\
End MJD & 55858.720 & 56233.501 \\
Nominal position & XIS & XIS \\ 
ObsID & 406046010 & 907005010 \\ 
\hline
XIS signal rate (cnt s$^{-1}$) & 0.18/0.18 & 5.1/4.5 \\
XIS exposure (ks) & 60.2 & 60.8 \\
PIN signal rate (cnt s$^{-1}$) & -- & 0.70  \\
PIN exposure (ks) & 53.7 & 50.5 \\
X-ray Intensity & 0.28 mCrab &  9.5mCrab \\
\hline
$F_{\rm x}$ ($10^{-12}$~erg~s$^{-1}$~cm$^{-2}$) 2--10 keV  &
 $5.36\pm0.03$ & 
 $190.0^{+0.2}_{-0.5}$
 \\
$F_{\rm x}$ ($10^{-12}$~erg~s$^{-1}$~cm$^{-2}$) 1--70 keV &
 $6.28^{+2.69}_{-0.52}$ & 
 $600.4^{+3.1}_{-4.1}$ \\
 \hline
$L_{\rm x}$ (erg s$^{-1}$)  1--70 keV  &
 $2.2\times 10^{33}$ &  $2.1\times 10^{35}$ 
\enddata
\tablecomments{The 0.5--10 keV XIS and 12--70 keV HXD-PIN signal rates are averaged during the observation. The XIS rates are shown for the XIS-FI and -BI CCD imagers. The source and background regions are stated in \S2.2. The X-ray intensity is evaluated in the 2--10 keV band. $F_{\rm x}$ and $L_{\rm x}$ are evaluated using the \texttt{comptt} model.}
\end{deluxetable}

\begin{deluxetable}{lcccccccc}
\tabletypesize{\footnotesize}
\tablecaption{
Comparison of the 2012 and 2011 time-averaged spectra of 4U~1954+319.
 \label{tab:TimeAveragedSpectra2012}}
\tablehead{
\colhead{Model} & \colhead{$N_{\rm H}$}	& 
\colhead{$\Gamma$} & \colhead{$E_{\rm cut}$} &
\colhead{$E_{\rm fold}$} & 
\colhead{$kT_e$} & 
\colhead{$kT$} & 
\colhead{$\tau$} & 
\colhead{$\chi^2$/dof (Prob.)} 
\\
\colhead{} & \colhead{($10^{22}$ cm$^{-2}$)} &
\colhead{} & \colhead{(keV)} & \colhead{(keV)} &
\colhead{(keV)} & 
\colhead{(keV)} & 
\colhead{} & 
\colhead{} 
}
\startdata
 \multicolumn{9}{c}{2012 AO7 ToO observation} \\
 \hline
 {\bf bbody+cutoffpl} & 
 $5.4\pm0.1$ & $1.21\pm0.01$ & --- & $19.9\pm0.5$ & --- & $0.13\pm0.01$ & --- &
 971.5/696($2.0\times 10^{-11}$)
 \\
  {\bf bbody+pow*highecut} & 
 $4.9\pm0.1$ & $1.24\pm 0.01$ & $4.23^{+0.17}_{-0.19}$ & $21.0^{+0.9}_{-0.8}$ & --- & 
 $0.13\pm0.01$ & --- & 
 890.1/696($6.9\times 10^{-7}$)
 \\
 {\bf bbody+NPEX} &  
 $4.9\pm0.1$ & $0.77\pm0.01$ & $6.6\pm0.1$ & --- & --- & $0.14\pm0.01$ & ---  &
 868.8/695($7.2\times 10^{-6}$)
 \\ 
 {\bf compTT} & 
 $1.7\pm0.2$ & --- & --- & --- & $7.5^{+0.2}_{-0.1}$ & $1.00\pm0.01$           & $8.9^{+0.13}_{-0.14}$
 & 780.8/697(0.015)
 \\ 
 \hline
 \multicolumn{9}{c}{2011 AO6 observation} \\
 {\bf bbodyrad} &
 $1.5\pm0.1$ & --- & --- & --- & --- & $1.28\pm0.01$ & --- &
 162.7/124(0.01)
 \\
 {\bf diskbb} &
 $3.0\pm0.1$ & --- & --- & --- & --- & $2.19\pm0.03$ & --- &
 162.3/124(0.12)
 \\
 {\bf cutoffpl} &
 $2.1^{+0.2}_{-0.1}$ & $-0.73\pm0.17$ & --- & $1.9\pm0.1$ & --- & --- & --- &
 133.9/123(0.23)
 \\
 {\bf pow*highecut} & 
 $2.0\pm0.2$ & $0.28^{+0.19}_{-0.20}$ & $3.0\pm0.1$ & $3.0\pm0.3$ & --- & --- & --- & 
 132.4/122(0.24)
 \\
  {\bf compTT} & 
 $1.6\pm0.1$ & --- & --- & --- & $<5.1$ & $0.85^{+0.10}_{-0.04}$  & $15.3^{+0.7}_{-8.8}$ 
 & 129.0/122(0.31) 
\enddata
\tablecomments{All the spectral uncertainties are given at 1$\sigma$ confidence level for single parameters. The photo-absorption and gaussian components are included in all the spectral models.
The parameter $E_{\rm cut}$ in the xspec \texttt{cutoffpl} model 
	is actually a folding energy, and shown in the $E_{\rm fold}$ column
	to compare the \texttt{highecut} model.}
\end{deluxetable}


\begin{deluxetable}{lcccccc}
\tabletypesize{\footnotesize}
\tablecaption{
Archival Suzaku X-ray pulsars comapred with 4U~1954+319
 \label{tab:SpectralComparison}}
\tablehead{
\colhead{X-ray source} &
\colhead{$P_{\rm s}$} & 
\colhead{$P_{\rm orb}$}  & 
\colhead{Optical} & 
\colhead{Obsid}	& 
\colhead{Exp} &
\colhead{Note} 
\\
\colhead{} &
\colhead{(sec)} &
\colhead{(day)} & 
\colhead{Type} & 
\colhead{} &
\colhead{(ks)} & 
\colhead{} 
}
\startdata
4U 1954+31       & 18300 & ??    & M4 III  & (Table. \ref{tab:Obslog}) & 60.2/60.8 & this work \\
GX 1+4           & 150   & 1161  & M5 III  & 405077010 & 97.2 & cutoffpl [1]  \\
4U 1700+24       & ??    & (404?)[2] & M2 III  & 402023010 & 50.3 & cutoffpl [3] \\
IGR J16194-2810  & ??    & ??    & M2 III  & 403024010 & 45.6 & cutoffpl [4] \\
\hline
Her~X-1          &  1.24 & 1.70 &  A9--B   & 100035010 & 36.9 & cutoffpl+cutoffpl [5] \\
1A~1118-616      & 408   & 24    & O9.5    & 403050010 & 22.0 & cutoffpl  \\
A~0535+262       & 103   & 111   & O9.7    & 404055010 & 31.8  & highecut \\
IGR J16393-4643  & 912   & 50.2 & B?[6][7]  & 404056010 & 50.6 & cutoffpl [8] \\
\hline 
4U 0114+65       & 9720  & 11.63 &  B1Ia   & 406017010 & 106.6  & cutoffpl+cutoffpl \\
4U 2206+54       & 5555  & 19.25 &  O9.5V  & 402069010 & 51.6 & cutoffpl \\
4U 1909+07       & 604   & 4.4   &  OB     & 405073010 & 29.2 & cutoffpl \\
\hline 
Aql X-1          & ---   & 19.0  &  K1     & 402053030 &  19.7 & bbody+compPS [9]
 \enddata
\tablecomments{
The photo-absorption is included in all the spectral models. Applied spectral models and references are noted. 
[1]\cite{Chakrabarty1997ApJ}
[2] 404 d period in 4U~1700+24 was not confirmed in \citep{Corbet2008ApJ}
[3]\cite{Masetti2006A&A, Nagae2008}
[4]\cite{Masetti2007A&A_J16194,Kitamura2013PASJ}
[5]\cite{Enoto2008PASJ}
[6]\cite{Bodaghee2012ApJ...751..113B} 
[7]\cite{Pearlman2011HEAD...12.4206P} 
[8]\cite{Nespoli2010A&A}
[9]\cite{Sakurai2012PASJ}
}
\end{deluxetable}


\clearpage


\end{document}